\documentclass[twocolumn,twocolappendix,trackchanges]{aastex631}

\newcommand{\pnt}[1]{{\scriptstyle#1}}
\usepackage{hyperref}
\usepackage{roboto}
\usepackage{threeparttable}
\usepackage{amsmath}
\usepackage{enumitem}
\usepackage{orcidlink}

\begin{document}

\title{Multiple emission lines of H$\alpha$ emitters at $z \sim 2.3$ from the broad and medium-band photometry in the ZFOURGE Survey}

\author{Nuo Chen \orcidlink{0000-0002-0486-5242}}
\affiliation{Department of Astronomy, Graduate School of Science, The University of Tokyo, 7-3-1 Hongo, Bunkyo-ku,
Tokyo 113-0033, Japan; \url{nuo.chen@grad.nao.ac.jp}}
\affiliation{National Astronomical Observatory of Japan, 2-21-1 Osawa, Mitaka, Tokyo 181-8588, Japan} 

\author{Kentaro Motohara \orcidlink{0000-0002-0724-9146}}
\affiliation{Department of Astronomy, Graduate School of Science, The University of Tokyo, 7-3-1 Hongo, Bunkyo-ku,
Tokyo 113-0033, Japan; \url{nuo.chen@grad.nao.ac.jp}}
\affiliation{National Astronomical Observatory of Japan, 2-21-1 Osawa, Mitaka, Tokyo 181-8588, Japan}
\affiliation{Department of Astronomical Science, SOKENDAI, 2-21-1 Osawa, Mitaka, Tokyo 181-8588, Japan}

\author{Lee Spitler \orcidlink{0000-0001-5185-9876}}
\affiliation{School of Mathematical and Physical Sciences, Macquarie University, Sydney, NSW 2109, Australia}
\affiliation{Macquarie University Astrophysics and Space Technologies Research Centre,  Macquarie University, Sydney, NSW 2109, Australia}
\affiliation{Australian Astronomical Optics, Faculty of Science and Engineering, Macquarie University, Macquarie Park, NSW 2113, Australia}

\author{Kimihiko Nakajima \orcidlink{0000-0003-2965-5070}}
\affiliation{National Astronomical Observatory of Japan, 2-21-1 Osawa, Mitaka, Tokyo 181-8588, Japan}

\author{Rieko Momose \orcidlink{0000-0002-8857-2905}}
\affiliation{Carnegie Observatories, 813 Santa Barbara Street, Pasadena, CA 91101, USA}

\author{Tadayuki Kodama \orcidlink{0000-0002-2993-1576}}
\affiliation{Astronomical Institute, Tohoku University, 6-3, Aramaki, Aoba, Sendai, Miyagi 980-8578, Japan}

\author{Masahiro Konishi \orcidlink{0000-0003-4907-1734}}
\affiliation{Institute of Astronomy, Graduate School of Science, The University of Tokyo, 2-21-1 Osawa, Mitaka, Tokyo 181-0015, Japan}

\author{Hidenori Takahashi}
\affiliation{Institute of Astronomy, Graduate School of Science, The University of Tokyo, 2-21-1 Osawa, Mitaka, Tokyo 181-0015, Japan}

\author{Kosuke Kushibiki}
\affiliation{Institute of Astronomy, Graduate School of Science, The University of Tokyo, 2-21-1 Osawa, Mitaka, Tokyo 181-0015, Japan}

\author{Yukihiro Kono}
\affiliation{Institute of Astronomy, Graduate School of Science, The University of Tokyo, 2-21-1 Osawa, Mitaka, Tokyo 181-0015, Japan}

\author{Yasunori Terao}
\affiliation{Institute of Astronomy, Graduate School of Science, The University of Tokyo, 2-21-1 Osawa, Mitaka, Tokyo 181-0015, Japan}

\begin{abstract}
We present a multiple emission lines study of $\sim$1300 H$\alpha$ emitters (HAEs) at $z \sim 2.3$ in the ZFOURGE survey. In contrast to the traditional spectroscopic method, our sample is selected based on the flux excess in the ZFOURGE-$K_s$ broad-band data relative to the best-fit stellar continuum. Using the same method, we also extract the strong diagnostic emission lines for these individual HAEs: [O{\sc iii}]$\lambda\lambda4959,5007$, [O{\sc ii}]$\lambda\lambda3726,3729$. Our measurements demonstrate good consistency with those obtained from spectroscopic surveys. 
We investigate the relationship between the equivalent widths ($EW$s) of these emission lines and various galaxy properties, including stellar mass, stellar age, star formation rate (SFR), specific SFR (sSFR), ionization states (O32). We have identified a discrepancy between  between HAEs at $z\sim2.3$ and typical local star-forming galaxy observed in the SDSS, suggesting the evolution of lower gas-phase metallicity ($Z$) and higher ionization parameters ($U$) with redshift. Notably, we have observed a significant number of low-mass HAEs exhibiting exceptionally high $EW_{\mathrm{[O\pnt{III}]}}$. Their galaxy properties are comparable to those of extreme objects, such as extreme O3 emitters (O3Es) and Ly$\alpha$ emitters (LAEs) at $z\simeq2-3$. Considering that these characteristics may indicate potential strong Lyman continuum (LyC) leakage, higher redshift anaglogs of the low-mass HAEs could be significant contributors to the cosmic reionization. Further investigations on this particular population are required to gain a clearer understanding of galaxy evolution and cosmic reionization.
\end{abstract}

\let\clearpage\relax

\section{Introduction} 
\label{sec:intro}
The rest-frame UV-optical-near-infrared spectrum of galaxies is characterized by a number of crucial emission lines originating from the interstellar medium (ISM) within them, including hydrogen recombination line and metal lines. These spectral lines serve as powerful indicators of the physical and chemical conditions in galaxies, providing insights into their stellar population, star formation rate (SFR), chemical abundance, and ionization properties \citep[e.g.,][]{Pagel79,Pagel92,Kewley02}.

Over the past decade, significant progress has been made in the study of gas-phase metallicities ($Z$) and ionization parameters ($U$) of galaxies, thanks to the use of emission line diagnostics. Notably, the ``Fundamental Metallicity Relation" (FMR), which explores the relationship between stellar mass ($M_*$), star formation rate (SFR), and metallicity, has been extensively investigated \citep[FMR; e.g.,][]{Mannucci10,Sanders18,Nakajima23}. This relation suggests that the accretion of pristine gas from the intergalactic medium (IGM) enhances the SFR while diluting the metallicity of the interstellar medium (ISM). \citet{Nakajima14} has used the O32 versus R23-index diagram to examine the ionization parameters of Ly$\alpha$ emitters at $z\sim2$. Their study revealed higher ionization parameters in these emitters compared to typical Lyman-break galaxies (LBGs) at similar redshifts. Building upon these findings, it is currently speculated that galaxies showing strong emission lines, such as [O{\sc iii}], may play a significant role in the cosmic reionization process \citep[e.g.,][]{Nakajima16,Yang17,Jaskot19,Tang19,Onodera20}.

Given the significance of emission line diagnostics, several extensive sky surveys have endeavored to extract these emission lines from large sets of observational data. Examples include the Sloan Digital Sky Survey \citep[SDSS;][]{York00,Kauffmann03} and the MOSFIRE Deep Evolution Field survey \citep[MOSDEF;][]{Kriek15}. Traditionally, spectroscopy has been the primary method used to analyze the spectra of galaxies. However, spectroscopic targets often encounter selection biases and have limitations in terms of the search volume, particularly for high-redshift galaxies. This can restrict the overall understanding of the galaxy population. Photometric observations with narrow-band (NB) filters also enable us to derive the emission line fluxes \citep[e.g.,][]{Sobral13}. However, because of the narrow redshift windows, NB imaging surveys needs comparably long observation time to construct large samples.

Encouragingly, recent studies have demonstrated the feasibility of identifying galaxies with strong emission lines using broad-band (BB) photometry \citep[e.g.,][]{Stark13, Saito20, Onodera20, Terao2022}. These studies have revealed that galaxies with high equivalent widths ($EW$s) of emission lines exhibit noticeable flux excess in broad-band photometry. For instance, \citet{Onodera20} utilized the COSMOS2015 catalog \citep{Laigle16} to select extreme [O{\sc iii}] emitters (O3Es) at $3<z<3.7$, based on the flux excess observed in the UltraVISTA-$K_s$ filter. Similarly, \citet{Terao2022} identified a substantial number of H$\alpha$ emiiters (HAEs) in the redshift range of $2.1<z<2.5$ using the ZFOURGE-$K_s$ filter. It is worth noting that these studies solely relied on a single broad-band photometric filter to search for emission line galaxies, without extracting emission line measurements from other photometric filters.

On the other hand, it is feasible to extract multiple emission lines information from multi-wavelength photometric data, particularly with the aid of medium-band (MB) photometry \citep[e.g.,][]{vanDokkum09}. Medium-band photometry employs narrower bandwidths ($\Delta\lambda/\lambda\sim15$) compared to broad-band filters, allowing for more precise wavelength sampling of emission lines. The ZFOURGE survey \citep{Straatman16} utilized the FourStar imager \citep{Persson13} to acquire near-infrared (NIR) medium-band photometry. Combined with other optical and infrared photometry, the ZFOURGE catalog offers a powerful tool for estimating multiple emission lines and conducting emission line diagnostics. By building composite spectral energy distributions (SEDs), \citet{Forrest18} categorized a significantly large galaxy sample at $1 < z < 4$ and characterized $\sim150$ extreme emission-line galaxies (EELGs, usually defined as $EW>300\mathrm{\AA}$). One notable advantage of employing photometry for multiple emission lines analysis is the ability to construct larger and more unbiased samples, which enhances the statistical significance of the results and allows for more robust conclusions. 

Currently, multiple emission lines analysis on galaxies at $z\sim2$ primarily relies on spectroscopic observations \citep[e.g.,][]{Reddy18, Topping20, Runco21}. These studies have investigated the connections between rest-frame optical emission lines and various physical attributes of galaxies, such as stellar mass, age, and SFR. Nonetheless, these spectroscopic studies have limitations on sample size and might inadvertently introduce a selection bias towards more massive and brighter galaxies. As a result, a comprehensive and unbiased sample is still lacking in this context. In this work, we address this gap by presenting a systematic search of H$\alpha$ emitters (HAEs) at $2.05 < z < 2.5$, mainly based on the photometric catalog from the ZFOURGE survey. By employing spectral energy distribution (SED) fitting with emission line templates, we construct a large sample of HAEs by measuring the H$\alpha$ emission line fluxes based on excesses observed in broad-band photometry. Furthermore, our method enables the simultaneous extraction of [O{\sc iii}] and [O{\sc ii}] emission lines from the flux excesses in the J and H medium-band photometry included in the ZFOURGE catalog. Our findings emphasize the feasibility of conducting emission line diagnostics solely based on photometry, which is particularly relevant in light of the ongoing large release of photometric data from the \emph{James Webb Space Telescope} (\emph{JWST}).

The outline of this paper is as follows. We introduce the ZFOURGE survey and other related dataset we used in this study in Section \hyperref[sec:obv]{2}. Sample selection, the SED fitting with the emission line templates and the basic measurements of line fluxes including H$\alpha$, [O{\sc iii}], [O{\sc ii}] are presented in Section \hyperref[sec:samp]{3}, and we derive physical parameters of the selected HAEs in the same Section. In Section \hyperref[sec:result]{4}, we carry out the mutiple emission lines analysis of HAEs and compare our sample with other analogous objects. In Section \hyperref[sec:dis]{5}, we point out the importance of low-mass HAEs found in our sample and discuss their relationship with galaxy evolution and the cosmic reionization. Finally we summarize our result in Section \hyperref[sec:conclu]{6} and present proposals for future observations to further investigate the properties of HAEs at $z\sim2.3$.

We adopt the following abbreviations for strong emission-line ratios:

\begin{equation}
    \mathrm{O32} = \mathrm{[O\pnt{III}]\lambda5007}\,/\,\mathrm{[O\pnt{II}]\lambda\lambda3726,29},
    \vspace{-0.3cm}
\end{equation}

\begin{equation}
    \mathrm{R23} = (\mathrm{[O\pnt{III}]\lambda\lambda4959,5007}+\mathrm{[O\pnt{II}]\lambda\lambda3726,29)}\,/\,\mathrm{H\beta}.
\end{equation}

Throughout this thesis, we adopt the AB magnitude system \citep{Oke83}, assume a Chabrier(\citeyear{Chabrier03}) initial mass function (IMF) and a $\mathrm{\Lambda CDM}$ cosmology with $H_0 = \mathrm{70\ km\ s^{-1}\ Mpc^{-1}}$, $\Omega_m = 0.3$, and $\Omega_{\Lambda} = 0.7$.

\section{Observations and data}
\label{sec:obv}
\subsection{The ZFOURGE survey}
\label{sec:zfsurvey}
We use the photometric catalog from the FourStar galaxy evolution survey \citep[ZFOURGE,][]{Straatman16}. ZFOURGE is a 45-night photometric observation survey with the FourStar near-infrared camera \citep{Persson13} on 6.5-meter Magellan telescope. The observation targets at three legacy fields: CDFS \citep{Giacconi02}, COSMOS \citep{Scoville07} and UDS \citep{Lawrence07} with a total coverage of $\sim450\,\mathrm{arcmin^2}$ (128, 135, 189 $\mathrm{arcmin^2}$ in CDFS, COSMOS, UDS, respectively).

The unique characteristic of ZFOURGE is that it has five near-infrared medium-band (MB) filters: $J_1$, $J_2$, $J_3$, $H_s$, $H_l$, covering a similar wavelength range as the broad-band (BB) filters $J$, $H$, and a ultra-deep $K_s$ map. ZFOURGE catalog also includes multiwavelength public data. In all, the CDFS, COSMOS, and UDS fields have 40, 37, and 26 photometric filters with the 80\% completeness of 26.0, 25.5, and 25.8 magnitudes in the $K_s$ images, respectively.

Such a large number of photometric filters makes it possible to accurately derive the photometric redshift (hereafter $z_{phot}$) of galaxies. \citet{Nanayakkara16} measure the spectroscopic redshift (hereafter $z_{spec}$) of $\sim200$ galaxies at $1.5 < z < 2.5$ in ZFOURGE-COSMOS and UDS field, and confirm that the primary $z_{phot}$ for star-forming galaxies (SFGs) from ZFOURGE catalog has a very good accuracy that $\Delta z/(1+z_{spec}) < 2\%$, where $\Delta z = |z_{spec}-z_{phot}|$.

\subsection{SWIMS medium K-band imaging}
\label{sec:swimsimage}
SWIMS \citep[Simultaneous-color Wide-field Infrared Multi-object Spectrograph,][]{Konishi12, Motohara16} is the first-generation near-infrared instrument for the University of Tokyo Atacama Observatory (TAO) 6.5m telescope \citep{Yoshii10}. 
It has medium K-band filters (detailed in Table \hyperref[tab:mediumk]{1}), which can provide more detailed information on the H$\alpha$ emission line at $z\sim2$. 
During its commissioning observation at the Subaru Telescope in S18B, an area of approximately $20\, \mathrm{arcmin}^2$ within the ZFOURGE-COSMOS Field has been observed, which contributes to nearly 1/6 of the total coverage of the ZFOURGE-COSMOS catalog. The total integration time is $\sim2$ hours for the $K_1$ filter and $\sim1.5$ hours for the $K_2$ filter, as outlined in Table \hyperref[tab:mediumk]{1}.

\begin{table}[t]
    \small
    \centering
    \label{tab:mediumk}
    \caption{SWIMS medium K-band and observation in S18B}
    \begin{tabular}{cccc}
        \hline\hline
        Filter & $\lambda$ & Depth$\mathrm{\,^{a}}$ & FWHM \\
        & $(\mathrm{\mu m})$ & (5$\sigma$, AB mag) & ($\farcs$) \\
        \hline
        $K_1$ & $1.95-2.09$ &  23.7 & $1\farcs0$\\
        $K_2$ & $2.10-2.24$ &  23.8 & $0\farcs6$\\
        \hline
    \end{tabular}
    \begin{tablenotes}
    \item \textbf{Notes.} ${ }^{\text {a }}$When calculating image depths (limiting magnitudes), we follow the same method as in \citet{Straatman16} and directly measure the fluxes of circular apertures with $0\farcs6$ diameter (same as ZFOURGE) placed at 5000 random positions on the final reduced images.
    \end{tablenotes}
\end{table}

The SWIMS data are reduced by a custom \robotoThin{Python-3} pipeline, named ``\robotoThin{SWSRED}'', which has demonstrated good stability and performance \citep{Konishi20}. To incorporate the SWIMS sources into the ZFOURGE catalog, we follow the same PSF matching method employed by the ZFOURGE survey (see Section 3.1 of \citet{Straatman16} for details). Among the objects at $z \sim 2.3$ in the ZFOURGE-COSMOS field, nearly 1/6 of them have detections in both the $K_1$ and $K_2$ filters and their medium K-band photometry is then merged into the ZFOURGE catalog. The SEDs are fitted as described below in Section \hyperref[sec:sed]{3.2}. The total integration times of the $K_1$/$K_2$ images are comparably shorter than those of the ZFOURGE ultra-deep $K_s$ images. Also, the coverage of SWIMS observation is much smaller than total ZFOURGE survey. Thus, we continue with the extraction of the H$\alpha$ emission line from the ZFOURGE $K_s$ filter in Section \hyperref[sec:extractline]{3.3}.

\section{Sample Selection and Measurements}
\subsection{Sample Selection}
\label{sec:samp}
\begin{figure}[t]
    \includegraphics[width=1\linewidth]{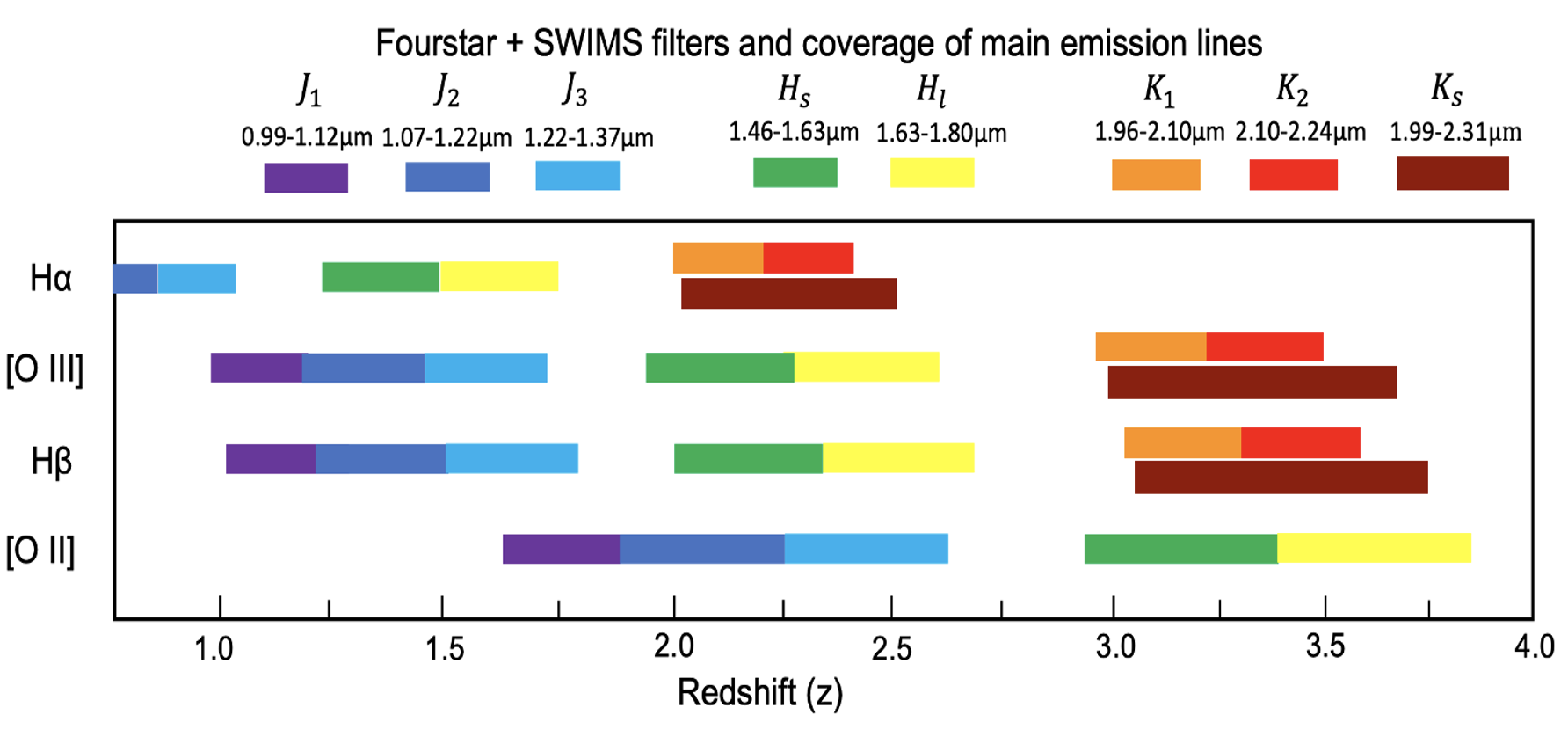}
    \label{fig:filter}
    \vspace{-0.5cm}
    \caption{Combinations of several strong emission lines, including H$\alpha$, [O{\sc iii}], H$\beta$ , [O{\sc ii}] in the observed-frame and the MB (BB) filters in which these emission lines drop. At $z \sim 2.3$, H$\alpha$ would fall into the K-band filters. Simultaneously, H$\beta$ and [O{\sc iii}] could be observed in the H medium-band filters and [O{\sc ii}] in the J medium-band filters.}
\end{figure}

We aim to construct a parent sample of galaxies with H$\alpha$ emission lines falling within the ZFOURGE $K_s$ filter. Based on the transmission curve of the $K_s$ filter, we determine that the H$\alpha$ emission line at $2.05 < z < 2.5$ is fully shifted into the $K_s$ filter as also shown in Figure \hyperref[fig:filter]{1}.

In the ZFOURGE catalog, $z_{phot}$ were obtained using the photometric reshift code, Easy and Accurate Z from Yale \citep[EAZY;][]{Brammer08}. Additionally, we merged the SWIMS $K_1$/$K_2$ fluxes into the ZFOURGE catalog. For objects that possess supplementary medium K-band data, we conducted a reiteration of the EAZY code and updated the $z_{phot}$ values with the newly generated outputs. The incorporation of the additional SWIMS MB data contributes to an enhanced constraint on the photometric redshifts (see in Appendix \hyperref[sec:swimsimageapx]{A}).

After updating the photometric redshifts in the ZFOURGE catalog, we compare them with the spectroscopic redshifts ($z_{spec}$) from the MOSDEF survey \citep{Kriek15} and the grism redshifts ($z_{grism}$) from the 3D-HST data release \citep{Brammer12, Momcheva16}. From the MOSDEF catalog, we have identified around 110 galaxies with high-quality spectroscopic redshift measurements at $z_{spec}\sim2.3$. These galaxies have also been cross-matched to the ZFOURGE-COSMOS catalog, allowing for a combined analysis of both datasets. The three legacy fields of ZFOURGE, namely COSMOS, UDS, and GOODS-S, are all covered by the 3D-HST survey. We have selected galaxies with high-quality $z_{grism}$ values (where \robotoThin{z\_best\_s} $\leq$ 2, please refer to \citet{Momcheva16} for more details) and cross-matched nearly 400 galaxies at $z_{grism}\sim2.3$ across the three fields. The majority of galaxies demonstrate a difference $\Delta z/(1+z) < 0.05$, with only 12 galaxies ($<3\%$) being outliers that have $\Delta z/(1+z) > 0.15$. To fully utilize the spectroscopic (grism) data, we proceed to replace the $z_{phot}$ values in the ZFOURGE catalog with corresponding $z_{spec}$ or $z_{grism}$ values obtained from the MOSDEF and 3D-HST Emission-Line Catalogs. In cases where both $z_{spec}$ and $z_{grism}$ are present, $z_{spec}$ takes precedence.

The ZFOURGE catalog employs a flag called \robotoThin{use}, which serves to eliminate various objects such as stars, objects in close proximity to stars, low signal-to-noise ratio (S/N) objects, and objects with low exposure time. We use this flag to eliminate contaminants that a standard selection of galaxies can be obtained by choosing sources with \robotoThin{use=1}. Additionally, the ZFOURGE catalog includes a list of AGN hosts that were identified through X-ray, IR, and radio selection methods, as outlined in \citet{Cowley16}. We include these AGNs in the following fitting, but exclude them when analyzing galaxy properties in Section \hyperref[sec:result]{4}. Applying these selection criteria, a total of 3754 galaxies at $2.05 < z < 2.5$ are retained, with 1307, 1235, and 1212 galaxies in the CDFS, COSMOS, and UDS fields, respectively.

\subsection{SED fitting with emission line templates}
\label{sec:sed}
In this study, we perform SED fitting to obtain primary galaxy properties using the 2020.0 version of Code for Investigating GALaxy Emission \citep[\robotoThin{CIGALE};][]{Burgarella05,Noll09,Boquien19}. We utilize the photometric data covering from $0.3-8\ \mathrm{\mu m}$ from ZFOURGE catalog and SWIMS.

The emission lines templates in \robotoThin{CIGALE} are computed based on user-defined gas-phase metallicities ($Z$) and ionization parameters ($U$), allowing for adjustable emission line templates during the SED fitting process. This is an update of SED fitting carried out in \citet{Terao2022} who have employed the same dataset but performed SED fitting using the Fitting and Assessment of Synthetic Templates \citep[\robotoThin{FAST};][]{Kriek09}. In addition, they have used a fixed emission-line template to be added into the spectrum, differing from ours where multiple parameter choices are available for emission line templates. Besides, \citet{Terao2022} solely considered the best fitting result from the template with the smallest $\chi^2$ value in their analysis, whereas we adopt a Bayesian-like approach, assigning weights to all models based on their $\chi^2$ values. This Bayesian fitting approach allows for obtaining a better estimate of the physical properties, such as stellar mass, with less uncertainties.

\subsubsection{Stellar population models and Star formation history}
\label{sec:sfh}
We utilize composite stellar population models generated from BC03 \citep{Bruzual03} with a Chabrier IMF \citep{Chabrier03}. The metallicity $Z$ of stellar population are permitted to be 0.004, 0.008 and 0.02. Next, we adopt a delayed-$\tau$ model to represent the star formation history (SFH) in a functional as follows:
\begin{equation}
\label{equ:delaytau}
    \mathrm{SFR}(t) \propto \frac{t}{\tau^{2}} \times \exp \left(-t / \tau\right)\quad\mathrm{for}\ 0 \leq t \leq t_0.
\end{equation}
The delayed-$\tau$ model offers a smooth SFH, characterized by an increases SFR from the onset of star-formation until it reaches its peak at $\tau$. Subsequently, the SFR gradually decreases. This model is considered to be a more representative SFH for SFGs compared to a constant SFH or an exponentially declining SFH \citep[e.g.,][]{Cohn18, Onodera20}. 

When establishing grids for our fitting process, we set the stellar population age ($t_0$) within the range of $\mathrm{log}(t_0/\mathrm{yr})=7–10$, with steps of 0.1 dex. The upper limit of $t_0$ is assumed not to exceed the age of the universe at $z \sim 2$. The e-folding time ($\tau$) ranges within $\mathrm{log}(\tau/\mathrm{yr})=8–10$, with steps of 0.1 dex.

\subsubsection{Nebular emission model}
\label{sec:emlinesed}
\robotoThin{CIGALE} models the emission of ionized gas in H{\sc ii} regions of the galaxy by using the nebular templates based on \citet{Inoue11} and implemented through \robotoThin{CLOUDY 13.0} \citep{Ferland98,Ferland13}. These nebular templates provide the relative intensities of 124 lines emitted by H{\sc ii} regions. The templates are parameterized according to a given ionization parameter $U$, and gas-phase metallicity $Z$ (which is assumed to be the same as the stellar one), along with a fixed electron density $n_e = 100\, \mathrm{cm}^{-3}$. In our fitting process, we consider an adjustable ionization parameter with values of $\mathrm{log\,U=-4 ,-3, -2, -1}$. A higher-resolution grid of $\mathbf{log\,U}$ have no effect on the results. Besides, Lyman continuum (LyC) photons are assumed to be completely absorbed by neutral hydrogen, i.e., $f_{esc} = 0$, and there is no LyC absorption by dust.

\subsubsection{Dust attenuation model}
\label{sec:dustext}
In this study, we fit the stellar continuum using the Calzetti curve supported by \robotoThin{CIGALE}. Also, stellar continuum and nebular emission usually suffer different dust extinction because H{\sc ii} regions possess a distinct distribution of dust or dust with different properties \citep{Calzetti94, Charlot00}. To address this issue, an approach is to assume that each component is subject to a different dust attenuation curve. For starburst galaxies, the Milky Way curve \citep{Cardelli89} is commonly adopted for the nebular line emission, even at higher redshift \citep[e.g.,][]{Reddy20}. Additionally, the extinction of the stellar continuum $E(B-V)_{star}$, and the extinction in the ionized gas $E(B-V)_{neb}$ are typically different. We parameterize the difference of color excesses by a factor $f$ such that:
\begin{equation}
\label{equ:ffactor}
    E(B-V)_{neb} = \frac{E(B-V)_{star}}{f}, \qquad (f < 1).
\end{equation}
While $f$-factor still suffers from significant uncertainty at high redshift \citep[e.g.,][]{Kashino13, Price14, Reddy20}, \citet{Saito20} proposed a simple redshift evolution for the $f$-factor as $f = 0.44 + 0.2z$. 
Thus, we adopt the Milky Way curve of \citet{Cardelli89} with an $f$-factor of 0.8 as the dust attenuation curves for the nebular emission.

\subsection{Emission line measurement}
\label{sec:extractline}
\begin{figure*}[hbt!]
    \centering
    \includegraphics[width=\textwidth]{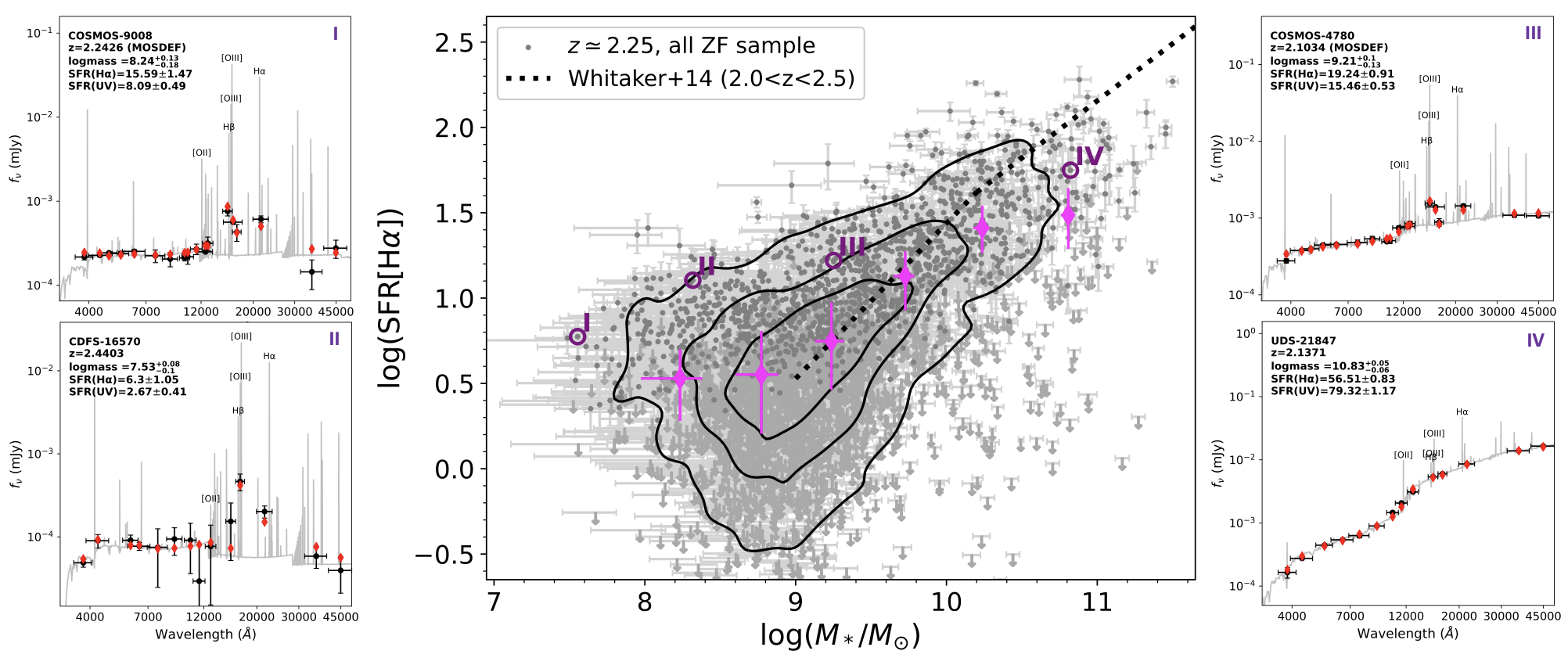}
    \label{fig:sfmsall}
    \vspace{-0.5cm}
    \caption{The star formation rate (H$\alpha$) as function of stellar mass, in the ZFOURGE fields. SFR(H$\alpha$) is derived from the calibration in \citet{Kennicutt12} (see Section \hyperref[sec:sfms]{3.6}). Grey circles show all the galaxies selected from flux excesses in $K_s$ photometry. 
    Each galaxy is given the error of stellar mass from SED fitting and the error of SFR($\mathrm{H\alpha}$) from $K_s$ photometry. Those $3\sigma$ upper limits for the $\mathrm{H\alpha}$-undetected sample have downward arrows. Magenta diamonds are median values from 6 mass bins with the median uncertainty on them. The best-fit $M_*$-SFR relation from \citet{Whitaker2014} are also shown as black dotted lines. Four example SEDs from the three ZFOURGE ﬁelds are given in the small panels. Among them, COSMOS-9008 have redshift measurement from MOSDEF \citep{Kriek15} and COSMOS-4780 have redshift measurement from 3D-HST \citep{Momcheva16}. Black circles are observed fluxes in several photometric ﬁlter sets and red diamonds are the best-fit SEDs convolved by the filter transmission curves. The grey spectrum is the best-fit SED based on CIGALE. Note that several optical median-band filter sets we used in the SED fitting are not shown here.}
    \vspace{0.5cm}
\end{figure*}

To obtain dependable emission line fluxes from the best-fit model and subsequently select emitters, we follow \citet{Terao2022} and employ the concept of ``flux excess" ($F_{excess}$), in units of $\mathrm{10^{-19}\,erg\,s^{-1}\,cm^{-2}}$. This value is calculated as the difference between the total observed flux and the flux of the stellar continuum derived from SED within a broad/medium-band filter of a bandwidth ($\Delta\lambda$).  The computation is as follows:
\begin{equation}
\label{equ:flexcess}
    F_{excess}\,(\mathrm{erg\,s^{-1}\,cm^{-2}}) = f_{obs} \times\Delta\lambda -  \int_{\lambda_1}^{\lambda_2} f_{cont}\,d\lambda,
\end{equation}
where $F_{excess}$ represents the total flux of all emission lines within a specific filter, theoretically being zero when no emission line falls within the filter, $\lambda_1$($\lambda_2$) the cut-off wavelength, $\Delta\lambda=\lambda_2-\lambda_1$ the bandwidth of the filter, $f_{obs}$ the observed flux density of the filter that the target emission line is located, and $f_{cont}$ the stellar continuum obtained from the best-fit SED model for each galaxy. The best-fit SED of CIGALE also returns the fluxes of the strong emission lines. By modeling different production rate of the Lyman continuum photons ($N_{\mathrm{lyc}}$) in a galaxy, \citet{Terao2022} have found that the derived $\mathrm{H\alpha}$ line flux from the emission line template strongly depends on the model assumption. On the other hand, the stellar continuum flux density from the SED fitting has been more robust against the model assumption. Besides, [O{\sc iii}] and [O{\sc ii}] emission line fluxes are strongly depend on the ISM properties ($Z$, $U$) assumed in the SED model, which cannot fully capture the true strengths of the emission lines.

One remaining issue in our method is the potential overlap of multiple emission lines within the same broad/medium-band filter due to its large bandwidth. It becomes challenging to isolate individual emission lines using only the broad/medium-band (BB/MB) data. To address this issue, we make an assumption and define a ratio, denoted as $r_{EL}$, which represents the target emission line's contribution relative to the combined strengths of all the emission lines present in the same filter.
\begin{equation}
    F_{EL}\,(\mathrm{erg\,s^{-1}\,cm^{-2}}) = r_{EL} \times F_{excess},
\end{equation}
where $F_{EL}$ is the final derived observed emission line fluxes and $F_{excess}$ is the flux excesses from Equation (\hyperref[equ:flexcess]{5}).
In Table \hyperref[tab:emission]{2}, we present the comprehensive information and criteria for each emission line used in our analysis.

\begin{table}[t]
    \small
    \centering
    \label{tab:emission}
    \caption{Main criteria for selecting galaxies at $z\sim2.3$}
    \begin{tabular}{ccc}
        \hline\hline
        Emission  & ZFOURGE Filter$\mathrm{\,^{a}}$ & Flux excess \\
        (continuum) &  &  (S/N) \\
        \hline
        $\mathrm{H\alpha}$ & $K_s$ & $3\sigma$  \\
        $\mathrm{[O\pnt{III}]}$ & $H_s$ or $H_l$ &  $2\sigma$ \\
        $\mathrm{[O\pnt{II}]}$ & $J_2$ or $J_3$  &  $2\sigma$\\
        \hline
    \end{tabular}
    \begin{tablenotes}
    \item \textbf{Notes.} ${ }^{\text {a }}$ The delicate design of $J$ and $H$-band medium filters make [O{\sc iii}] and [O{\sc ii}] emission lines drop into $H_s$ and $J_2$ filters at $z < 2.258$, and into $H_l$ and $J_3$ filters at $z > 2.258$, simultaneously. See also in Figure \hyperref[fig:filter]{1}.
    \end{tablenotes}
\end{table}

\subsubsection{$\mathrm{H\alpha}$}
\label{sec:extractha}
In the $K_s$ filter, the main contaminants include [N{\sc ii}]$\lambda\lambda6548,84$ and [S{\sc ii}]$\lambda\lambda6717,31$. We refer to the MOSDEF Emission-Line Catalog \citep{Kriek15} to estimate typical emission-line ratios at $2.05 < z < 2.5$. Initially, we remove galaxies that show non-detection of [N{\sc ii}], [S{\sc ii}] and $\mathrm{[N{\pnt{II}]\lambda6584\,/\,H\alpha}} > 0.5$ in the MOSDEF catalog. The later criterion is introduced because such strong [N{\sc ii}] emission is unlikely to be associated with star formation and could indicate the presence of AGN hosts \citep[BPT diagram;][]{Kauffmann03}. This selection yields a sample of 453 objects. From this sample, we obtain the line ratios, H$\alpha$/(H$\alpha$+[N{\sc ii}]+[S{\sc ii}]) and take their average to obtain $0.67\pm0.10$. Finally, we adopt $r_{\mathrm{H\alpha}} = 0.7$. To further validate this assumption, we also compute the emission line ratio from the best-fit SED of the entire HAE sample, and the resulting median value yields $r_{\mathrm{H\alpha}} \simeq 0.78$. Thus, we believe that adopting $r_{\mathrm{H\alpha}} = 0.7$ introduces only minor systematic errors. 
From the model spectrum, we could also obtain the contamination ratio of each galaxy. However, the contamination ratios from best-fit model spectra show almost no correlation to the spectroscopic measurements.

[S{\sc ii}] falls outside the $K_s$-band for 121 (9\%) galaxies at $2.45 < z < 2.5$, for which the contamination ratios $r_{\mathrm{H\alpha}}$ become larger. However, because of the $\sim2\%$ uncertainty of $z_{phot}$ in our study, we cannot identify which galaxies are exactly falling within the redshift range above. Thus, we apply a constant $r_{\mathrm{H\alpha}}$ regardless of the $z_{phot}$ of the galaxies.

To identify $\mathrm{H\alpha}$ emitters in the $K_s$-band filter, we implement a selection process based on flux excesses and photometric errors. 
We select candidates of HAEs by requiring the flux excesses in the $K_s$-band exceed three times the photometric errors ($>3\sigma$), that is,
\begin{equation}
    F_{excess,K_s}  >  3\times\Delta f_{K_s}\times \Delta\lambda.
\end{equation}
This criterion yield a sample of 1358 $\mathrm{H\alpha}$ emitters (439, 481, 438 in CDFS, COSMOS, UDS, respectively) at $z_{med} = 2.25$. In Figure \hyperref[fig:sfmsall]{2}, we display the complete sample of galaxies from the ZFOURGE catalog on the $M_*$-SFR diagram, including galaxies with flux excesses below $3\sigma$. Additionally, we show four examples of galaxies with their corresponding best-fit SEDs. Notably, the proximity between the observed flux and the model flux reinforces the robustness of our fitting methodology. The two low-mass examples show a distinct characteristic on the rest-frame optical to near-IR bands, displaying a flat continuum. This feature signifies the presence of young stellar populations. Also, these two galaxies exhibit a noticeable excess flux in the $K_s$-band, which is boosted by the strong $\mathrm{H\alpha}$ emission line.

\begin{figure}[hbt!]
    \includegraphics[width=1\linewidth]{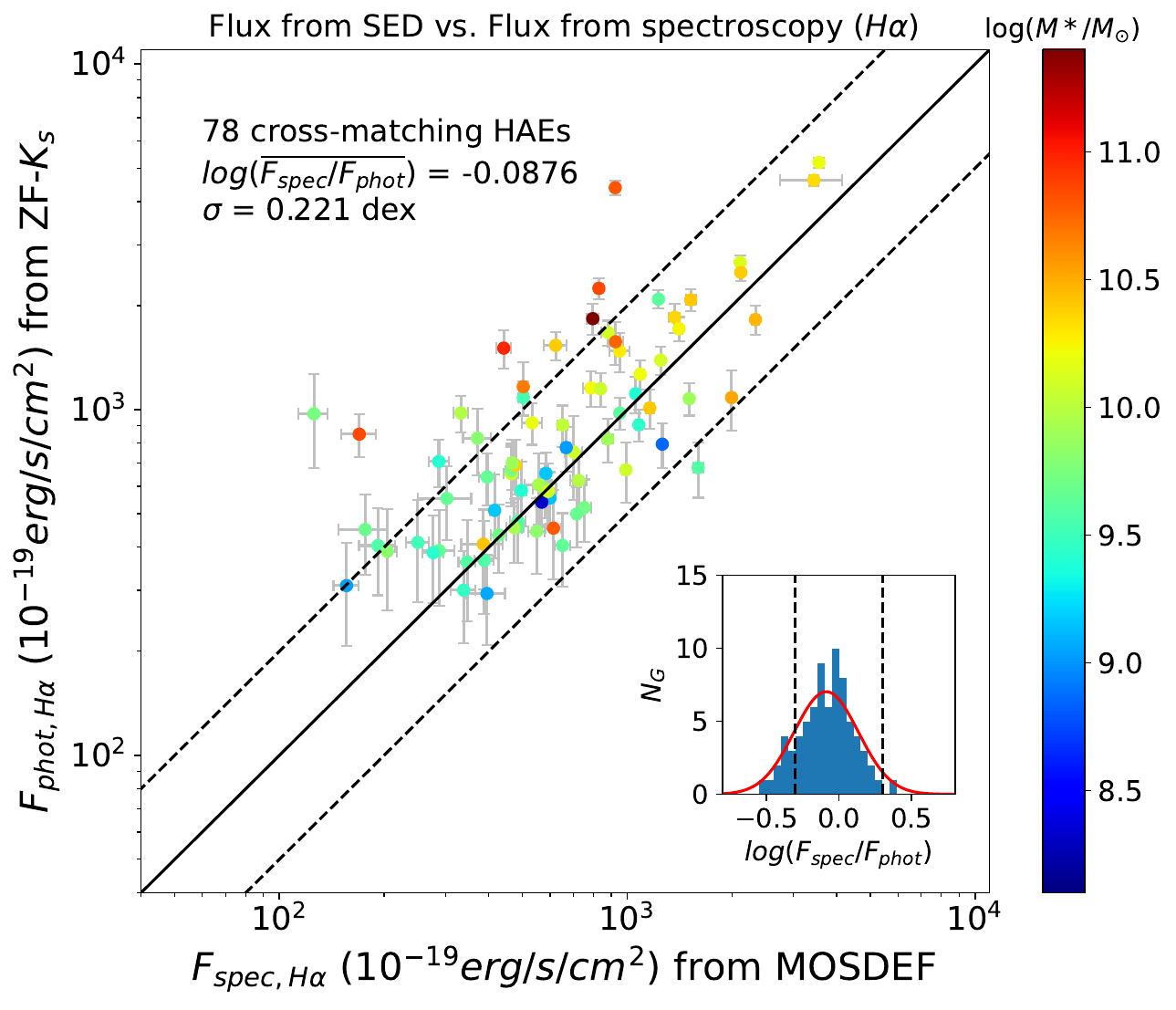}
    \label{fig:hamosdef}
    \vspace{-0.4cm}
    \caption{Comparison between the observed $\mathrm{H\alpha}$ fluxes derived from SED fitting and those from the MOSDEF spectroscopic emission-line catalog \citep{Kriek15}. Black dashed line indicates an agreement with a factor of 2. The error bars on the $y$-axis are the flux errors from the MOSDEF catalog, while those on the $x$-axis is from our method. Galaxies are separated into histograms on the lower right panel according to their residual from a 1:1 line. Here, the flux ratios ($F_{spec}/F_{phot}$) are scaled to $\mathrm{log_{10}}$-space with steps of 0.05 and black dashed lines (a factor of 2) are added. The mean difference and scatter on this one-to-one relation are also added. The color gradient of dots shows the stellar mass of individual galaxy. The overall estimation of emission line fluxes agrees well with the spectroscopic measurements, indicating the robustness of our method.}
\end{figure}

In order to validate the reliability of the emission lines obtained from the SED fitting results, we perform a comparison between the observed $\mathrm{H\alpha}$ fluxes ($F_{\mathrm{H\alpha}}$) in our work and the slit-loss-corrected fluxes obtained from the MOSDEF Emission Line Catalog \citep[$F$\textsubscript{spec},][]{Kriek15}. \citet{Reddy15} has introduced the slit-loss corrections of the MOSDEF survey by modeling the HST light profile of each galaxy, resulting in a silt-loss within 18\%. This comparison is conducted for a total of 78 galaxies in the ZFOURGE-COSMOS field, all of which have $\mathrm{H\alpha}$ detections with an S/N $>$ 3, according to both the MOSDEF catalog and our method. Figure \hyperref[fig:hamosdef]{3} presents the comparison of $\mathrm{H\alpha}$ fluxes between our method and the MOSDEF catalog. We find that 63 out of the 78 (81\%) of $\mathrm{H\alpha}$ emitters exhibit consistent flux values within a factor of 2, demonstrating agreement between the two datasets. This analysis further strengthens the confidence in the emission line measurements derived from our SED fitting approach.

\subsubsection{$\mathrm{[O\pnt{III}]}$ and $\mathrm{[O\pnt{II}]}$}
\label{sec:o3o2}
\begin{figure*}[hbt!]
    \includegraphics[width=0.49\textwidth]{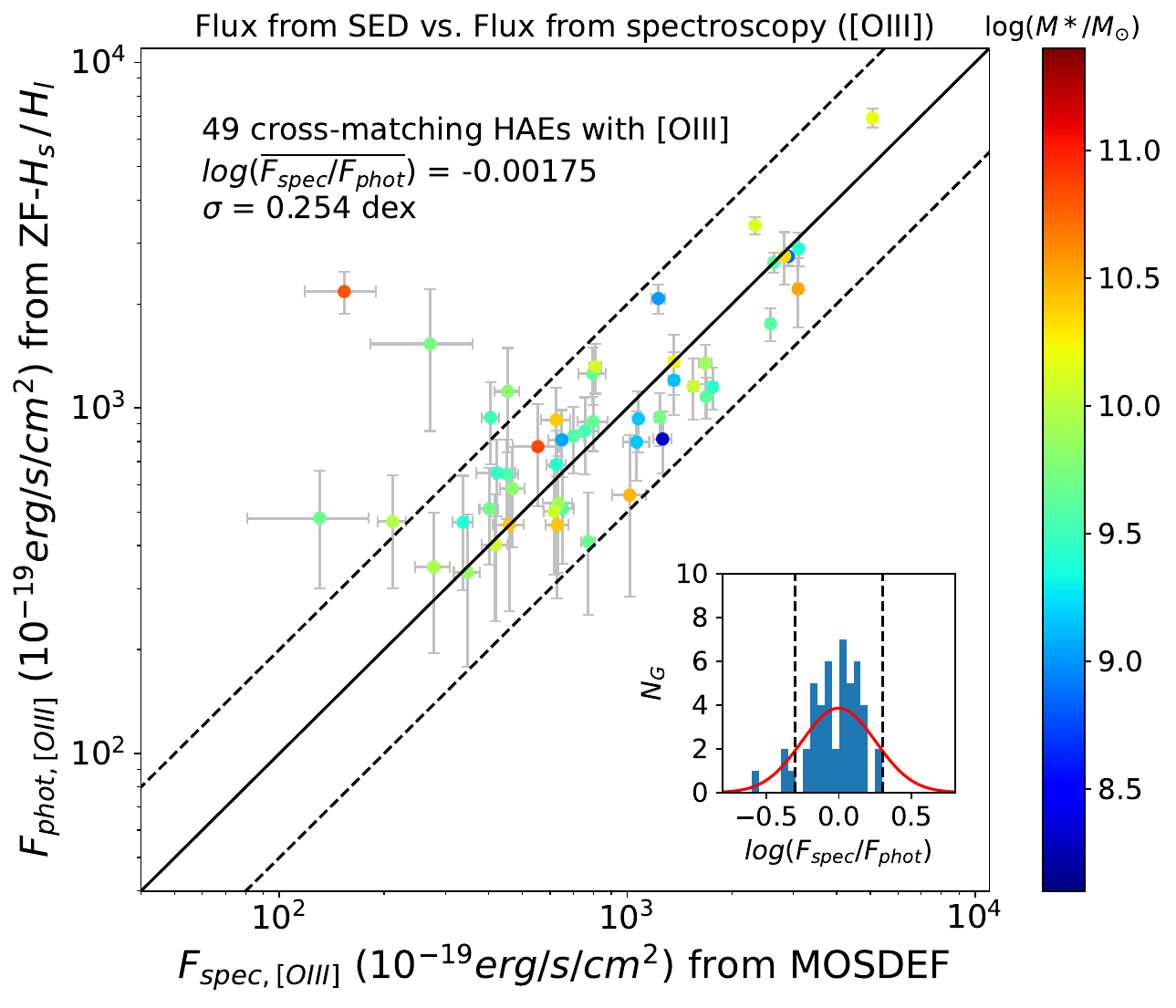}
    \includegraphics[width=0.49\textwidth]{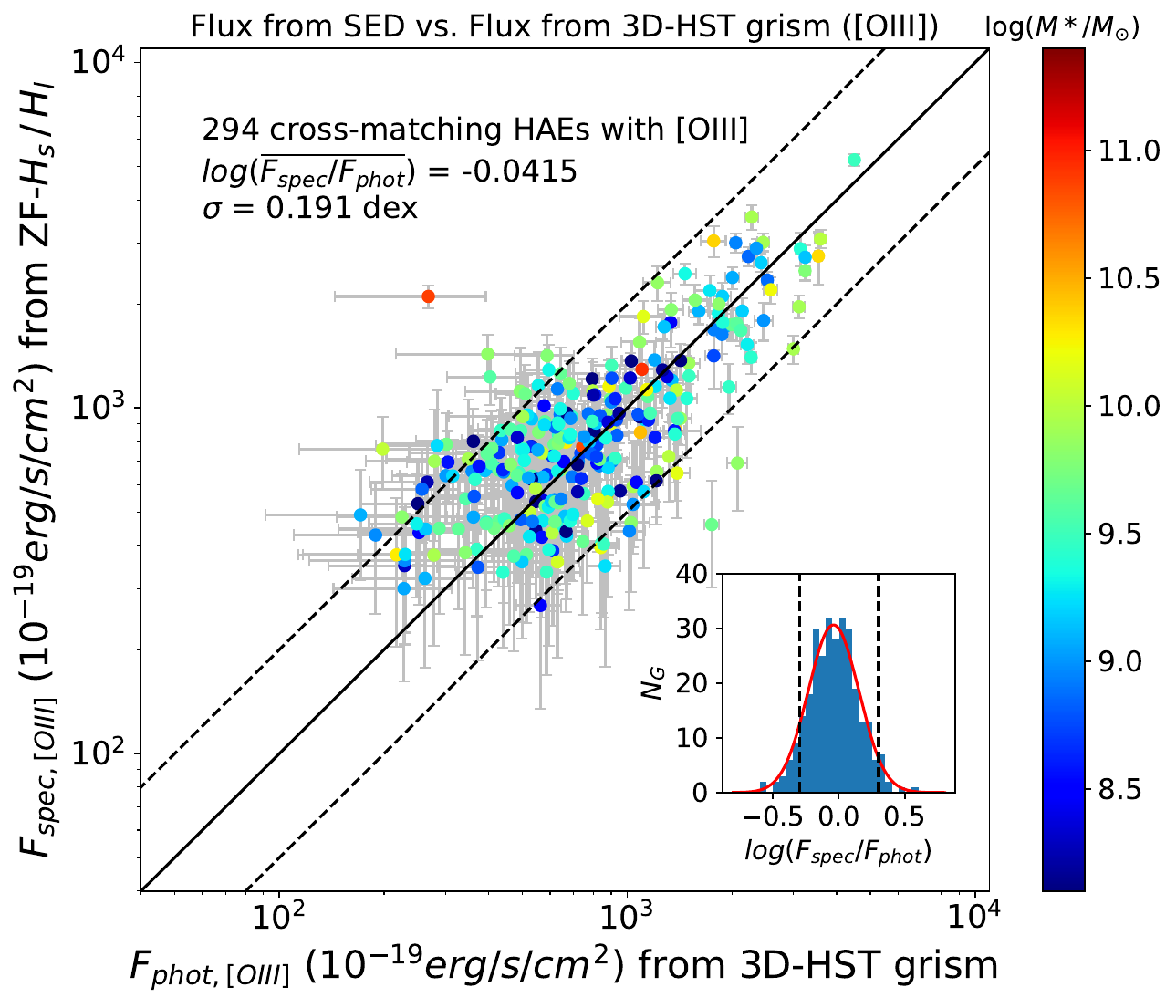}
    \label{fig:O3mosdef}
    \caption{Left: Comparison between the observed [O{\sc iii}]$\lambda\lambda4959,5007$ fluxes derived from SED fitting and those from the MOSDEF spectroscopic emission-line catalog \citep{Kriek15}. Right: Same for the observed [O{\sc iii}] fluxes but comparing with the grism spectra catalog from the 3D-HST survey \citep{Momcheva16}. Plot details as in Figure \hyperref[fig:hamosdef]{3}. The estimates agree, suggests our method works, even at low masses.}
    \vspace{0.5cm}
\end{figure*}

\begin{figure*}[hbt!]
    \includegraphics[width=0.49\textwidth]{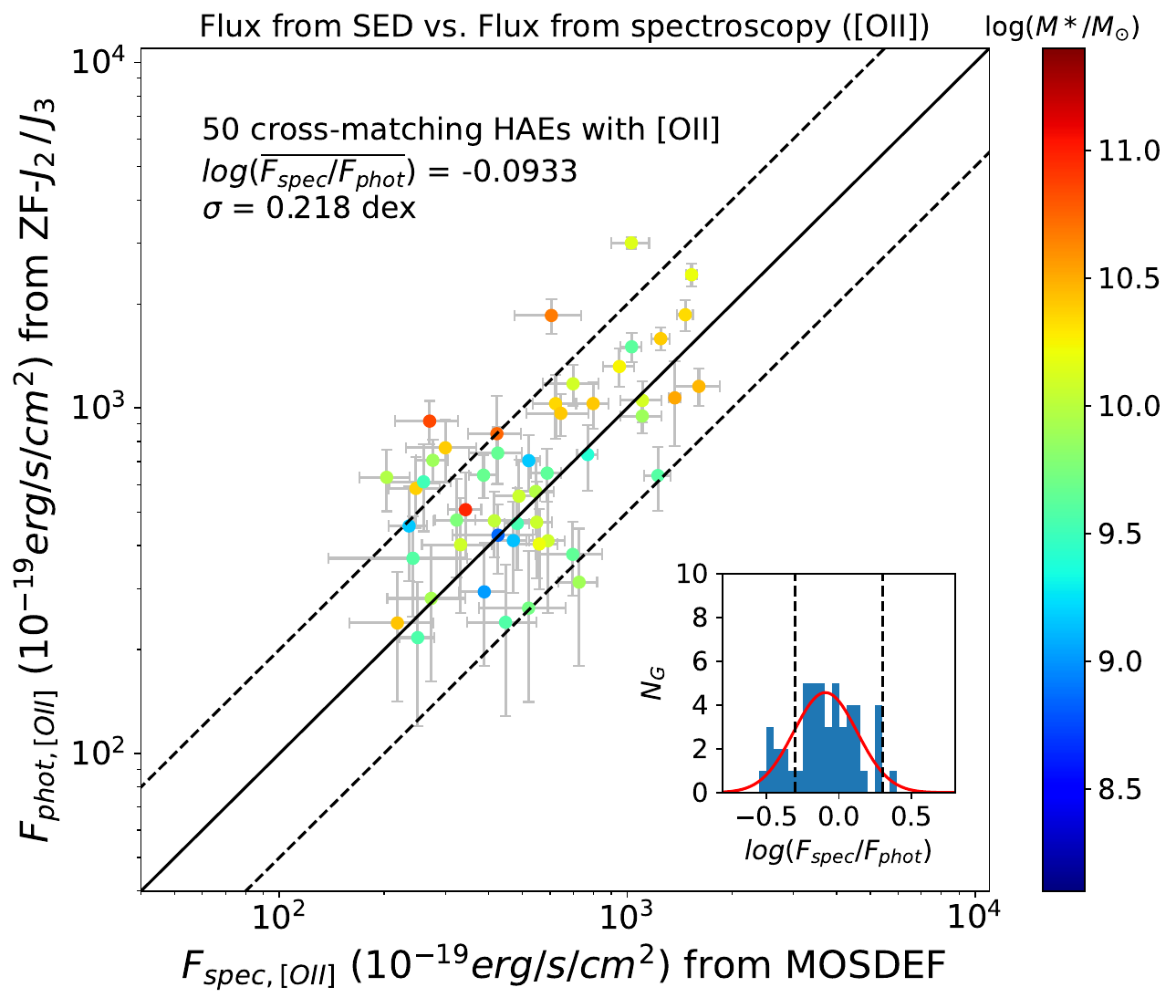}
    \includegraphics[width=0.49\textwidth]{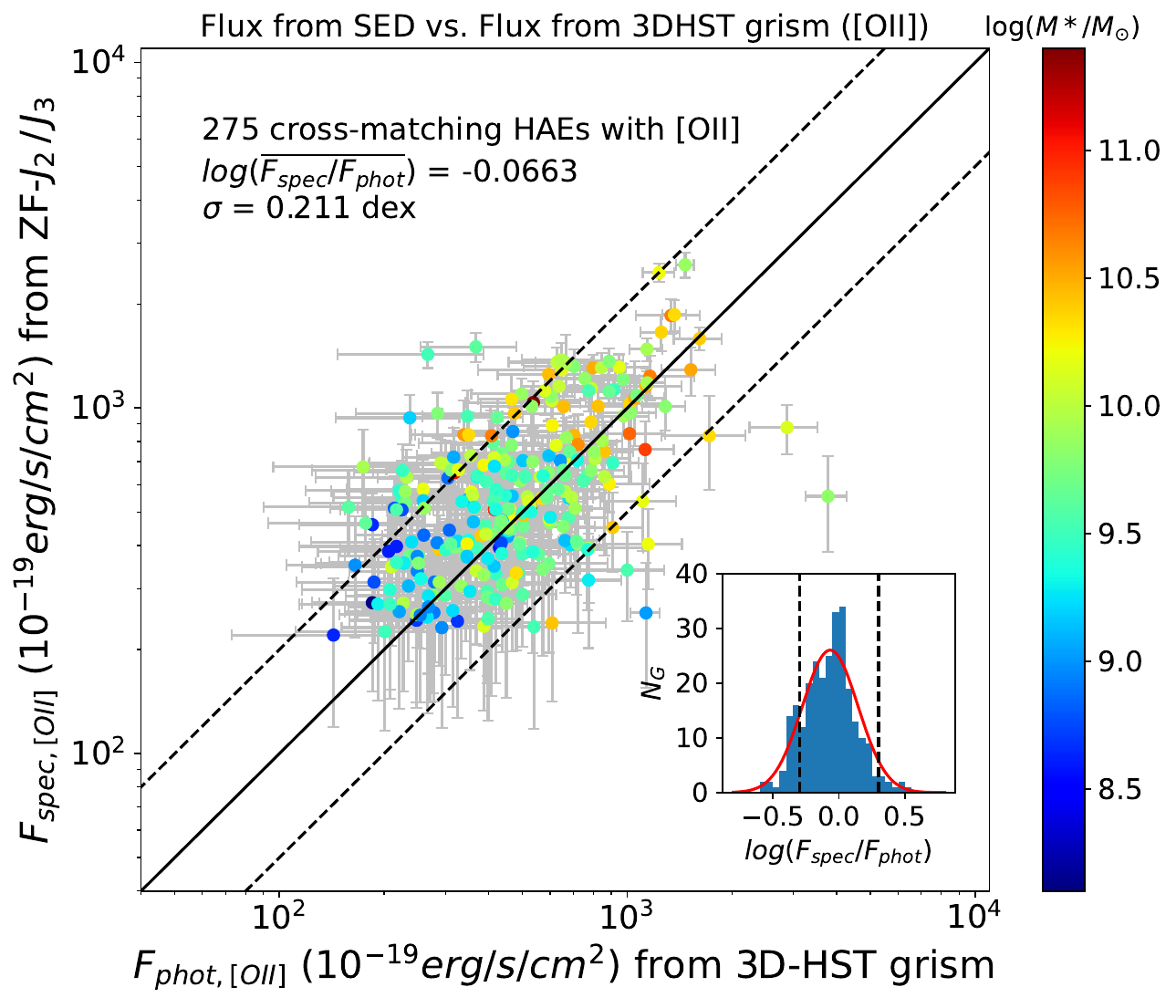}
    \label{fig:O2mosdef}
    \caption{Left: Comparison between the observed [O{\sc ii}] fluxes derived from SED fitting and those from the MOSDEF Emission-Line Catalog \citep{Kriek15}. Right: Same for the observed [O{\sc ii}] fluxes but comparing with the grism spectra catalog from the 3D-HST survey \citep{Momcheva16}. Plot details as in Figure \hyperref[fig:hamosdef]{3}. Again, agreement indicates our method is robust.}
    \vspace{0.5cm}
\end{figure*}

For galaxies at $z\sim2.3$, [O{\sc iii}] emission lines would drop in either the $H_s$/$H_l$ filter. To ensure accurate measurements of [O{\sc iii}] line fluxes, we assume that the total flux excesses are contaminated by H$\beta$. Building upon this, we adopt a Case-B recombination with $T_e=10,000K$ and $n_e = 100\, \mathrm{cm}^{-3}$. This allows us to derive the intrinsic H$\beta$ fluxes from the intrinsic H$\alpha$ fluxes using the following relation, 
\begin{equation}
\label{equ:balmerdecre}
    F_{\mathrm{H\beta},int}   = \frac{F_{\mathrm{H\alpha},int}}{2.86}.
\end{equation}
Here, $F_{\mathrm{H\alpha},int}$ and $F_{\mathrm{H\beta},int}$ are corrected for dust extinction from the observed fluxes $F_{\mathrm{H\alpha},obs}$ and $F_{\mathrm{H\beta},obs}$. 
By subtracting the observed flux
$F_{\mathrm{H\beta},obs}$ from the total flux excesses, we can obtain the [O{\sc iii}] emission line fluxes.

H$\beta$ falls outside the $H$-band medium filter for 221 (16\%) galaxies at $2.27 < z < 2.33$. Again, due to the $z_{phot}$ uncertainties, we do not make adjustment for the H$\beta$ contamination fraction depending on the redshift of the galaxies.

We expect [O{\sc ii}] emission lines at $z\sim2.3$ to be detected in either the $J_2$/$J_3$ filters. We estimate the contamination in a similar manner as for $\mathrm{H\alpha}$. The main sources of contamination include [Ne{\sc iii}]$\lambda\lambda3870,3969$ and Balmer lines such as H$\varepsilon\,\lambda3970$. Likewise, we refer to the MOSDEF catalog to obtain the average contamination ratio and set $r_{\mathrm{[OII]}} = 0.7$. Again, the contamination ratios from best-fit model spectra show no correlation to the spectroscopic measurements.

The contamination lines falls outside the $J$-band medium filter for galaxies at $2.45 < z < 2.5$, which is same to the case of H$\alpha$. We still apply a constant $r_{\mathrm{[OII]}}$ when deriving the [O{\sc ii}] emission.

To demonstrate the reliability of the emission line fluxes, we also compare the [O{\sc iii}] and [O{\sc ii}] emission line fluxes with those obtained from the MOSDEF catalog and 3D-HST catalog. Figure \hyperref[fig:O3mosdef]{4} and Figure \hyperref[fig:O2mosdef]{5} present these comparisons, revealing a consistency with the spectroscopic measurements.

In conclusion, our derivation of the $\mathrm{H\alpha}$, [O{\sc iii}], [O{\sc ii}] fluxes demonstrate no significant systematic biases (within 0.1 dex offset) and small scatter within 0.3 dex with those obtained from spectroscopic surveys, particularly for galaxies with relatively low masses down to the limits of the MOSDEF and 3D-HST survey. This suggests that the emission line fluxes estimated based on the flux excesses are robust and reliable.

\subsection{Rest-Frame Equivalent Widths}
Rest-frame $EW_\mathrm{{H\alpha}}$, $EW_\mathrm{{[O\pnt{III}]}}$ and $EW_\mathrm{{[O\pnt{II}]}}$ are calculated by dividing the line flux by the continuum flux density at a certain rest-frame wavelength. The continuum flux density is determined using the best-fit SED model through the following process. Firstly, we exclude the flux density points that contain both continuum and emission from the model. Next, we fit the continuum flux density points to a power-law slope, $f_v\propto\lambda^{\alpha}$ within the wavelength windows of $(\lambda_0-100)\times(1+z)$ to $(\lambda_0+100)\times(1+z)$, where $\lambda_0$ represents the rest-frame wavelength of the emission line in Angstrom. Finally, the fitted continuum at $\lambda_0\times(1+z)$ is
taken as the desired continuum flux density. Note that in Section \hyperref[sec:result]{4}, $EW_\mathrm{{[O\pnt{III}]}}$ refers to the combined equivalent widths of the [O{\sc iii}]$\lambda\lambda4959,5007$ doublet.

\subsection{Emission Line Ratios}
Based on our measurements, the available emission line diagnostics in our study are $\mathrm{O32}$ and $\mathrm{R23}$. Note that, we derive the intrinsic H$\beta$ fluxes from the intrinsic H$\alpha$ fluxes, as explained in Section \hyperref[sec:o3o2]{3.3.2}. To obtain the intrinsic $\mathrm{O32}$ and $\mathrm{R23}$ values, the related emission lines are corrected for dust attenuation, $E(B-V)_{neb}$, which is obtained from the Bayesian SED fitting result. 

Note that, the definition of [O{\sc iii}] in these two diagnostics is slightly different; $\mathrm{O32}$ uses [O{\sc iii}]$\lambda5007$, while $\mathrm{R23}$ uses [O{\sc iii}]$\lambda\lambda4959,5007$. We use line ratio of $\mathrm{[O\pnt{III}]}\lambda5007:\mathrm{[O\pnt{III}]}\lambda4959=2.97:1$ for the conversion.

\subsection{Star Formation Main Sequence}
\label{sec:sfms}
In this study, we quantify star formation rates (SFRs) in galaxies using the $\mathrm{H\alpha}$ indicator, converted by the calibration in \citet{Kennicutt12}, with a correction applied for the \citet{Chabrier03} IMF: 
\begin{equation}
    \mathrm{log\,SFR}(\mathrm{H} \alpha) = \mathrm{log}\,L_\mathrm{H\alpha} - 41.34
\end{equation}
The amount of dust attenuation is measured through dust extinction $E(B-V)_{neb}$ obtained from the Bayesian SED fitting result.

Generally, star-forming galaxies have a correlation between stellar mass ($M_*$) and SFR, known as the ``Main Sequence" (SFMS), which holds true at least up to $z \sim 3$ \citep[e.g.,][]{Whitaker2014,Speagle2014}. However, intriguingly, several recent studies based on observations of low-mass galaxies at both low and high redshift have revealed that their samples show elevated SFRs relative to the main sequence \citep[e.g.,][]{Hayashi2016, Onodera20, Terao2022, Atek22}.

\begin{figure}[t]
    \centering
    \includegraphics[width=1\linewidth]{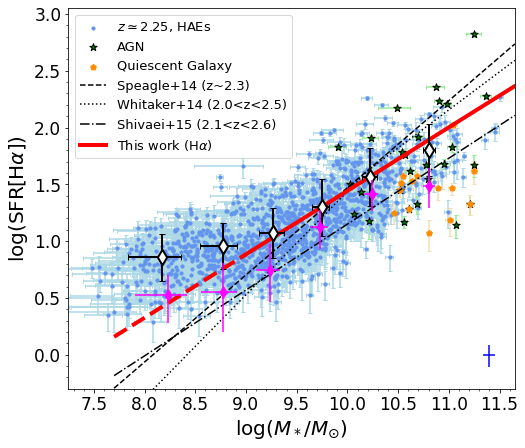}
    \label{fig:sfms}
    \vspace{-0.3cm}
    \caption{The star formation main sequence (SFMS) of 1318 HAEs at $z_{med} = 2.25$ in the ZFOURGE fields. HAEs are required to have $>3\sigma$ flux excesses in the $K_s$ filter. Red solid line is the best linear fit to the galaxies with $\mathrm{log}(M_*/M_{\odot})>9.0$, which is extrapolated to lower mass with red dashed line. The best-fit SFMS from \citet{Whitaker2014}, \citet{Speagle2014} and \citet{Shivaei15} are also shown with black dotted, dashed, and dot-dashed lines, respectively. Active galactic nuclei (AGNs) and quiescent galaxies (QGs) are marked as green stars and orange pentagons respectively. These are excluded from the fitting of SFMS. White diamonds are the median values from 6 mass bins with the median uncertainty on the HAEs. While, magenta diamonds in Figure \hyperref[fig:sfmsall]{2} are also added here for comparison. It is obvious that those low-mass HAEs ($<10^9\,M_{\odot}$) tend to scatter above the SFMS($\mathrm{H\alpha}$).}
\end{figure}

In Figure \hyperref[fig:sfms]{6}, we present the SFMS between $\mathrm{SFR(H\alpha)}$ and the stellar mass for the 1318 HAEs in our catalog. To achieve a full completeness in fitting the SFMS of our sample, we consider the mass completeness limit of the ZFOURGE survey, which is $\mathrm{log}(M_*/M_{\odot})\sim9$ at $z\sim2$ \citep{Straatman16}. At the high-mass end, we exclude the AGN identified in \citet{Cowley16} and quiescent galaxies selected using the $UVJ$ diagram \citep[e.g.,][]{Wuyts07, Spitler12} from our fitting. This results in a total of 40 AGNs and quiescent galaxies being excluded. Applying linear regression fitting, we obtain the following relationship between $\mathrm{SFR(H\alpha)}$ and $M_*$,
\begin{equation}
\label{equ:sfms}
\begin{aligned}
    \mathrm{log\,SFR}(\mathrm{H} \alpha) = & (0.56\pm0.03) \times \log M_* \\  & -(4.15\pm0.28).
\end{aligned}
\end{equation}

The slope in Equation \hyperref[equ:sfms]{10} is different from that in \citet{Whitaker2014},  which has a slope of 0.91 at the low-mass end. \citet{Shivaei15} has reported that the slope of the SFMS can be influenced by various observational and measurement factors. These factors can contribute to the discrepancies in the slopes reported in different studies, as shown in Figure \hyperref[fig:sfms]{6}. In our study, the sample biases primarily arise from the selection criterion of HAEs. In Figure \hyperref[fig:sfmsall]{2}, all galaxies are included without the $3\sigma$ requirement, where we observe a good agreement between the data points and the extrapolated SFMS from \citet{Whitaker2014}. Additionally, if we apply a $2\sigma$ criteria for the HAEs selection (in Table \hyperref[tab:emission]{2}), the resulting SFMS slope is found to be $0.69\pm0.03$. These observations suggest that the selection criterion and its associated sample biases have a notable impact on the derived slope of the SFMS, emphasizing the need to carefully account for such effects when interpreting and comparing results across different studies.\\

We have compiled a catalog of $\mathrm{H\alpha}$ emitters for the three ZFOURGE fields. The catalog includes a total of 1318 HAEs with redshifts ranging from 2.05 to 2.5. Each entry in the catalog provides information such as coordinates, observed emission line fluxes, flux uncertainties, and SED-derived properties including stellar mass, stellar age, and dust attenuation. The identification of individual sources in the catalog is based on their unique ID, which corresponds to the ZFOURGE catalog.
Within the full sample of HAEs, 859 sources have a detection of [O{\sc iii}] emission lines, while 824 sources have a detection of [O{\sc ii}] emission lines. Additionally, there are 626 HAEs that have both [O{\sc iii}] and [O{\sc ii}] emission lines detected. This catalog provides a comprehensive dataset for studying HAEs at $z\sim2.3$.

\section{Mutiple Emission Lines Analysis of H$\alpha$ emitters}
\label{sec:result}
In this section, we investigate the relationship between rest-frame equivalent widths and various properties of the HAEs at $z\sim2.3$. Specifically, we examine their dependence on stellar mass, stellar age, star formation rate (SFR), and specific star formation rate (sSFR). Additionally, we explore the correlations between equivalent widths and the available emission line index, O32. For comparison, we also consider other galaxy samples from different studies. This includes star-forming and star-burst ($EW_\mathrm{{H\alpha}}>50\mathrm{\AA}$) galaxies from the SDSS DR7 MPA/JHU catalog \citep{Kauffmann03, Brinchmann04}, extreme O3Es at $z\sim2.2$ \citep{Tang19}, galaxies from the MOSDEF survey in the mid-redshift windows at $z\sim2.3$ \citep{Kriek15,Reddy18}. By comparing the properties and trends among these different samples, we aim to gain a comprehensive understanding of the relationships of these attributes.

Given the substantial sample size, we perform a best-fit linear correlation analysis and report the Spearman's rank correlation coefficient, $r_s$, of these variables in Table \hyperref[tab:ewproper]{3}. The Spearman statistical test underscores the robustness of the correlation between equivalent widths (log[$EW/\,\mathrm{\AA}$]) and the physical parameters of galaxies.

\subsection{$EW$ vs. Stellar properties}
\label{sec:ewsfr}
\subsubsection{$\mathrm{[O\pnt{III}]}$ EWs}
\begin{figure*}
    \centering
    \includegraphics[width=0.49\textwidth]{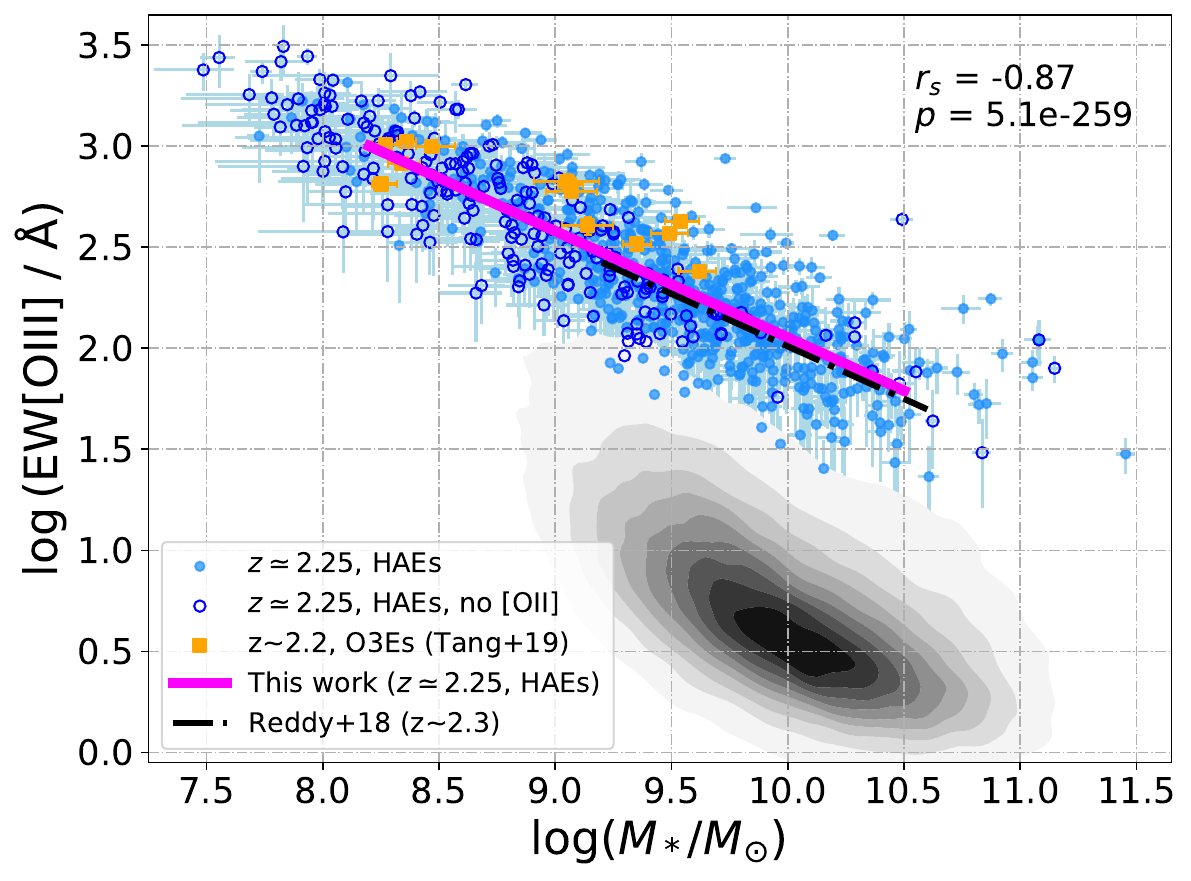}
    \includegraphics[width=0.49\textwidth]{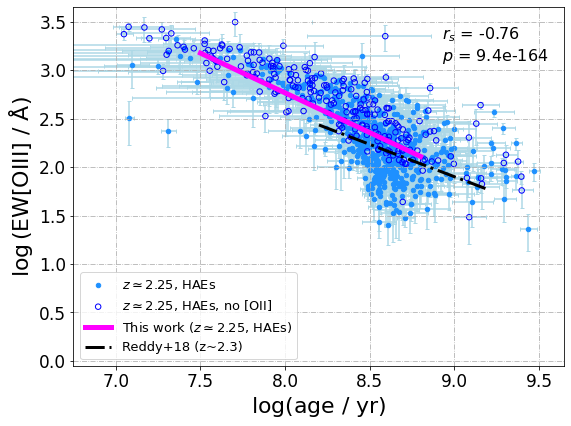}
    \includegraphics[width=0.49\textwidth]{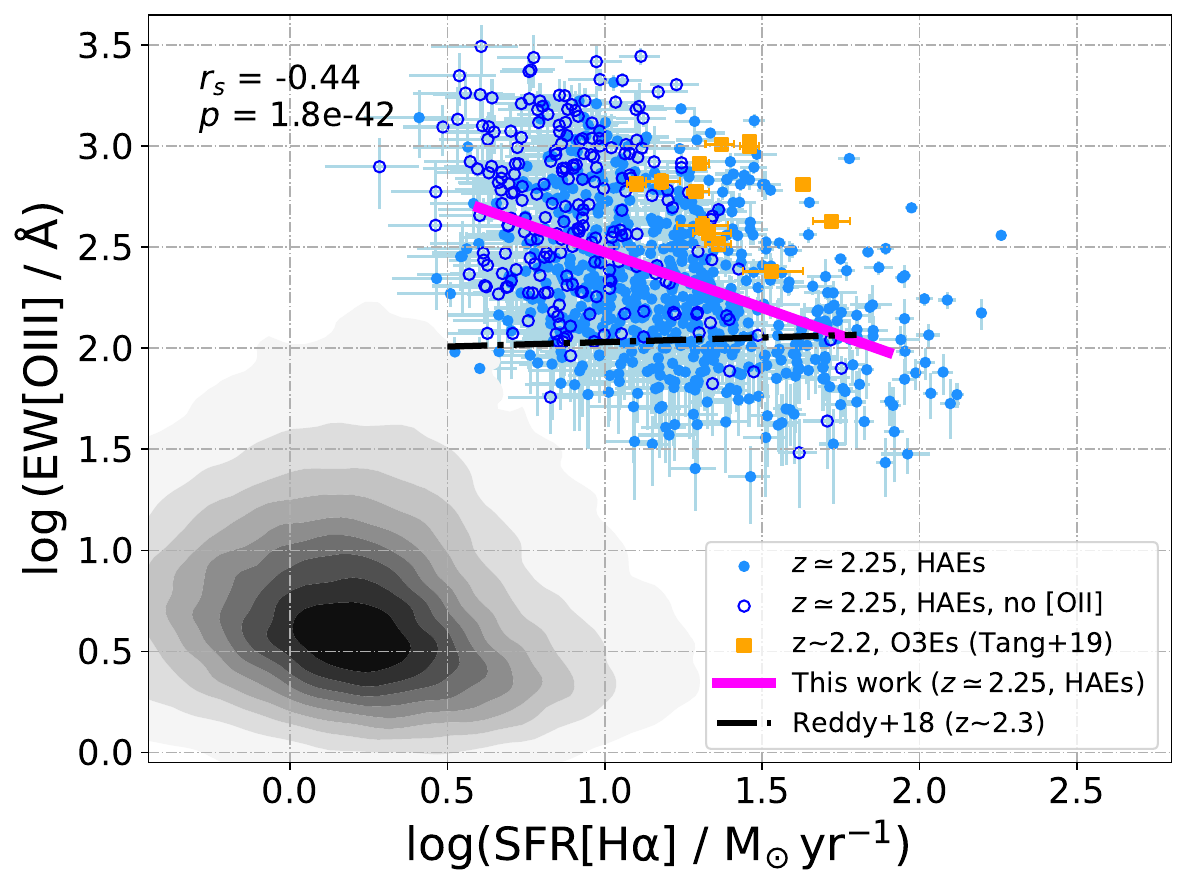}
    \includegraphics[width=0.49\textwidth]{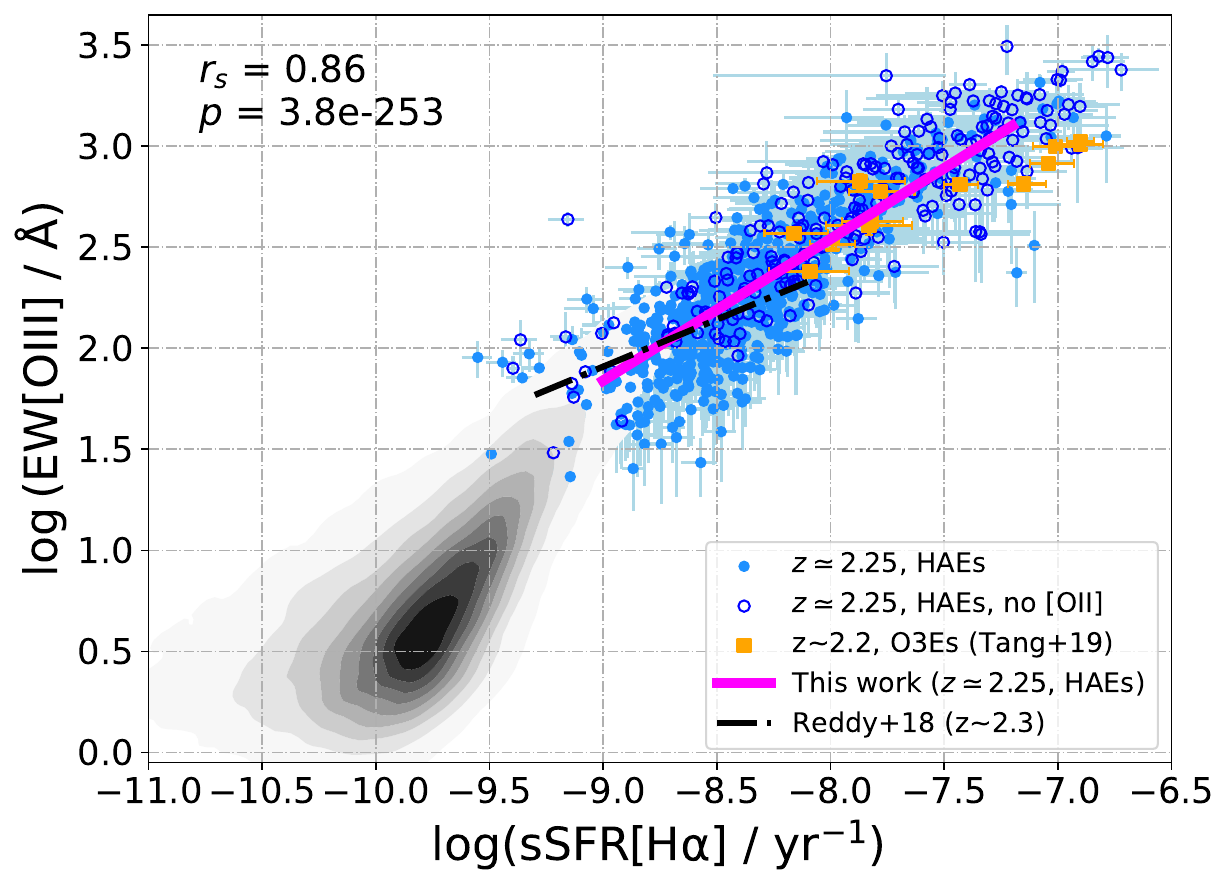}
    \label{fig:ewo3vsgal}
    \caption{Relationship between the $EW_\mathrm{{[O\pnt{III}]}}$ and stellar properties (stellar mass, age, SFR, sSFR) of the 859 HAEs at $z\sim2.3$ in our study. The stellar mass and stellar age of each sample is derived from the Bayesian result of \robotoThin{CIGALE}. The upper left, the upper right, the lower left, and the lower right panel shows the stellar mass, the stellar age, the SFR, the sSFR, versus $EW_\mathrm{{[O\pnt{III}]}}$, respectively. In each panel, the magenta solid line tracks the best-fit linear correlation of the full sample. Those HAEs with both [O{\sc iii}] and [O{\sc ii}] detection are marked as blue solid circles, while HAEs with only [O{\sc iii}] detection are presented by blue open circles. For comparison, the SDSS star-forming/star-burst galaxies are represented by contours, with the densest region depicted in black. Besides, extreme O3Es at $z\sim2.2$ from \citep{Tang19} are marked as orange square in these panels. The red dashed-dotted line present the relationship for $z\sim2.3$ massive galaxies from MOSDEF \citep{Reddy18}. $EW_\mathrm{{[O\pnt{III}]}}$ of our sample exhibits a strong correlation to stellar mass, age and sSFR.}
    \vspace{0.5cm}
\end{figure*}

\begin{figure*}
    \centering
    \includegraphics[width=0.49\textwidth]{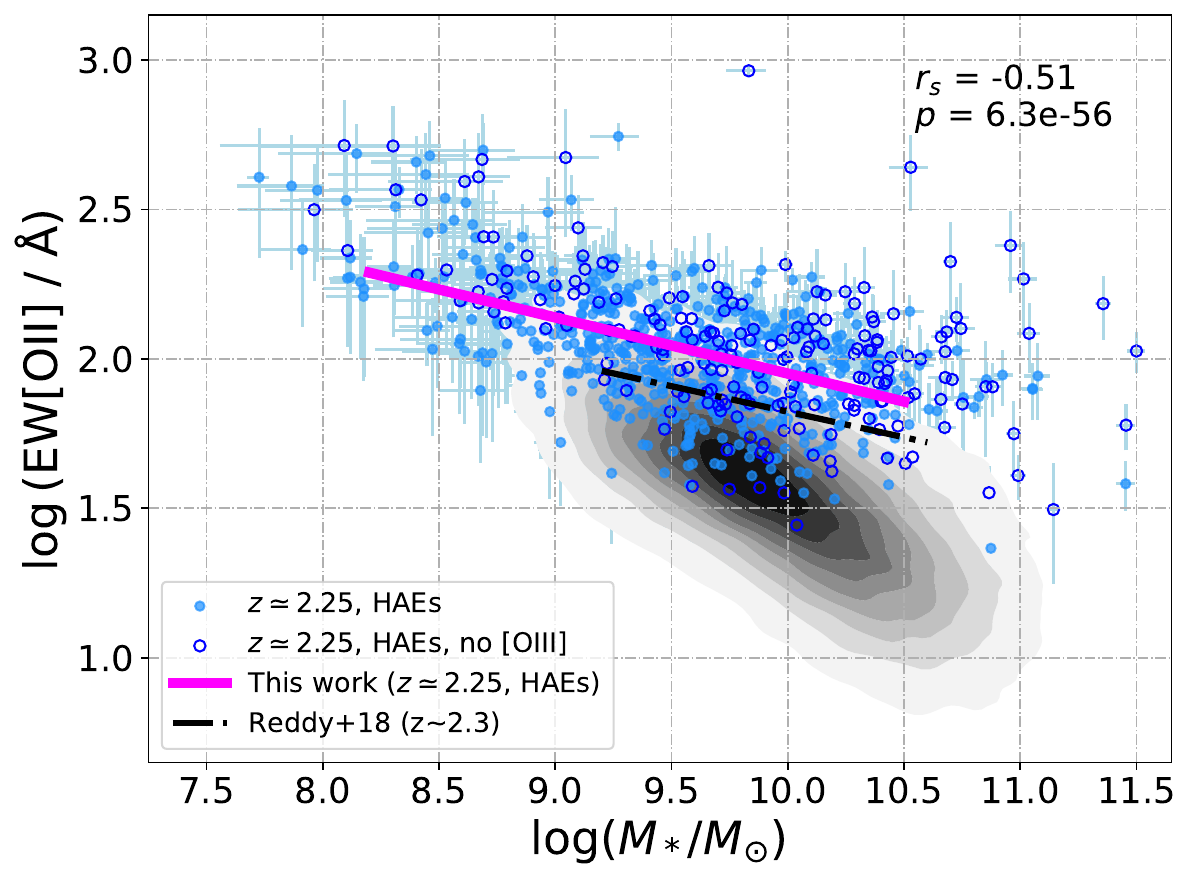}
    \includegraphics[width=0.49\textwidth]{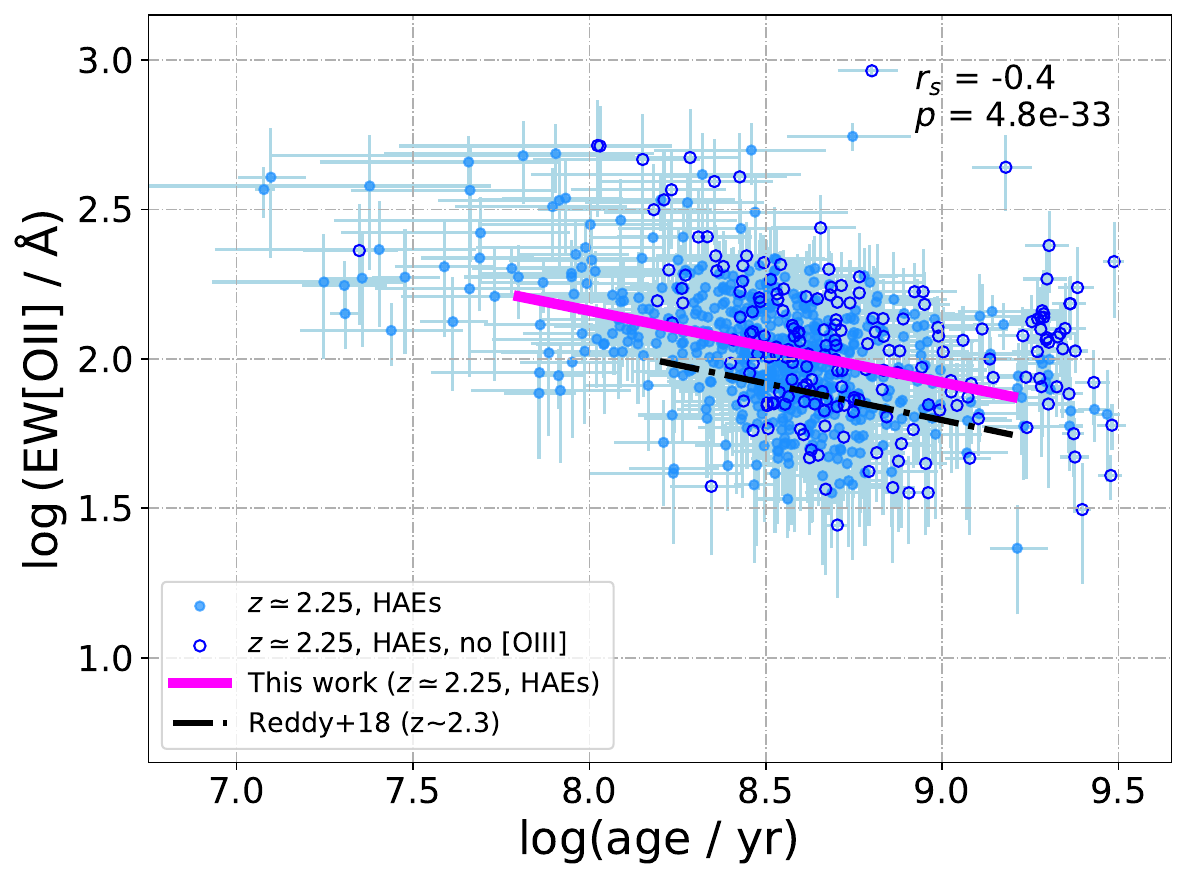}
    \includegraphics[width=0.49\textwidth]{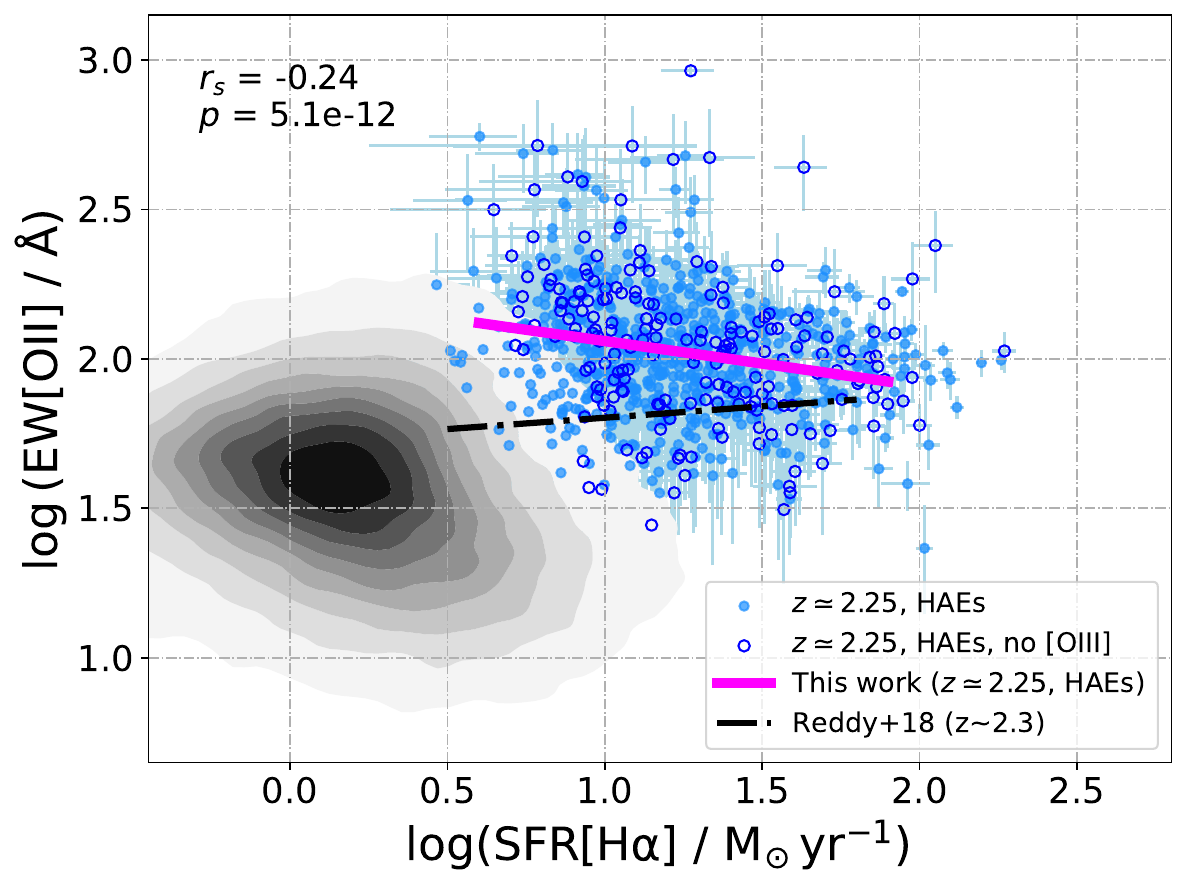}
    \includegraphics[width=0.49\textwidth]{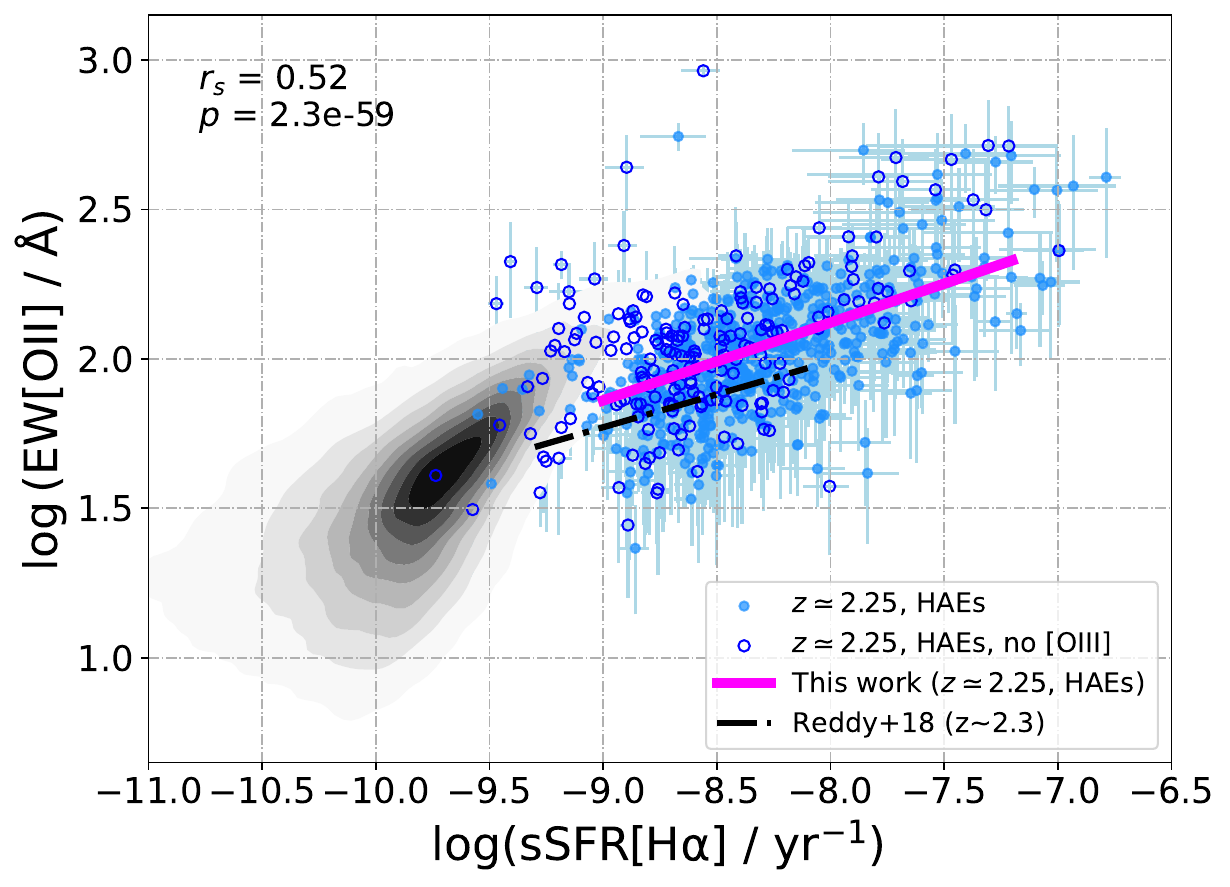}
    \label{fig:ewo2vsgal}
    \caption{Relationship between the $EW_\mathrm{{[O\pnt{II}]}}$ and stellar properties (stellar mass, age, SFR, sSFR) of the 824 HAEs with [O{\sc ii}] at $z\sim2.3$ in our study. Outlines as in Figure \hyperref[fig:ewo3vsgal]{7}. Here, those HAEs with both [O{\sc iii}] and [O{\sc ii}] detection are marked as blue solid circles, while HAEs with only [O{\sc ii}] detection are presented by blue open circles.}
    \vspace{0.5cm}
\end{figure*}

\begin{figure*}
    \centering
    \includegraphics[width=0.49\textwidth]{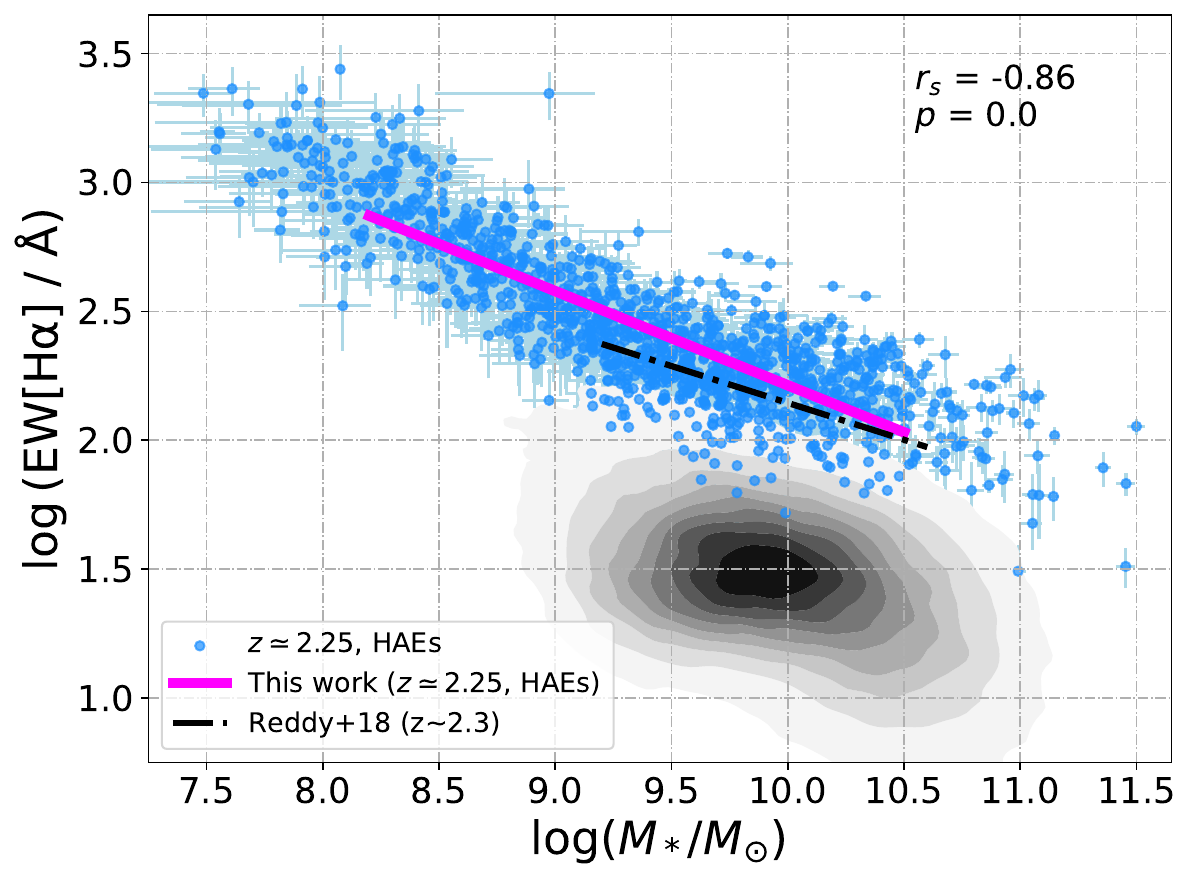}
    \includegraphics[width=0.49\textwidth]{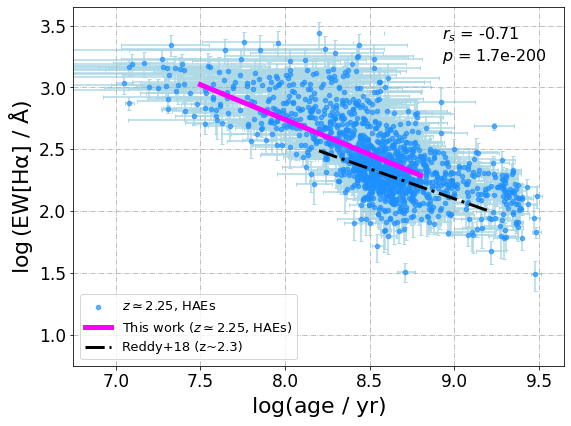}
    \includegraphics[width=0.49\textwidth]{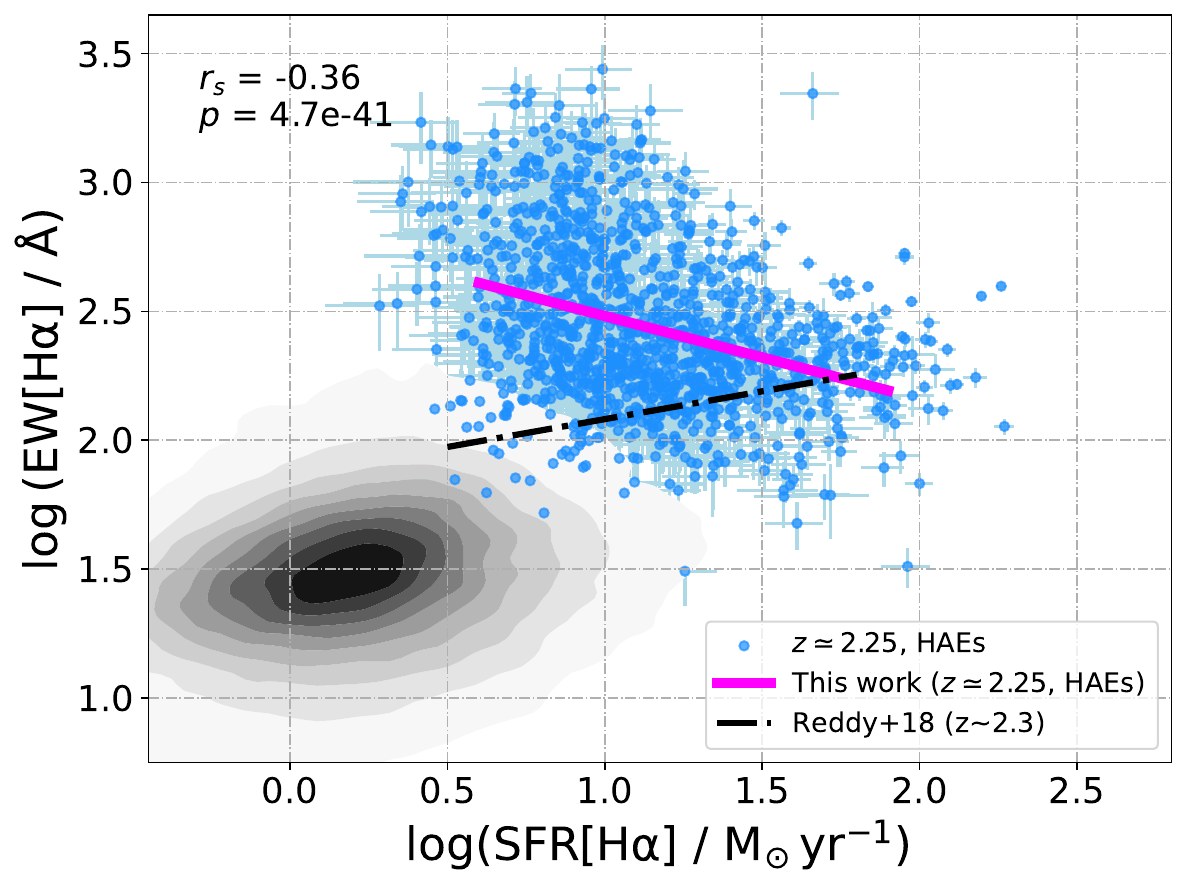}
    \includegraphics[width=0.49\textwidth]{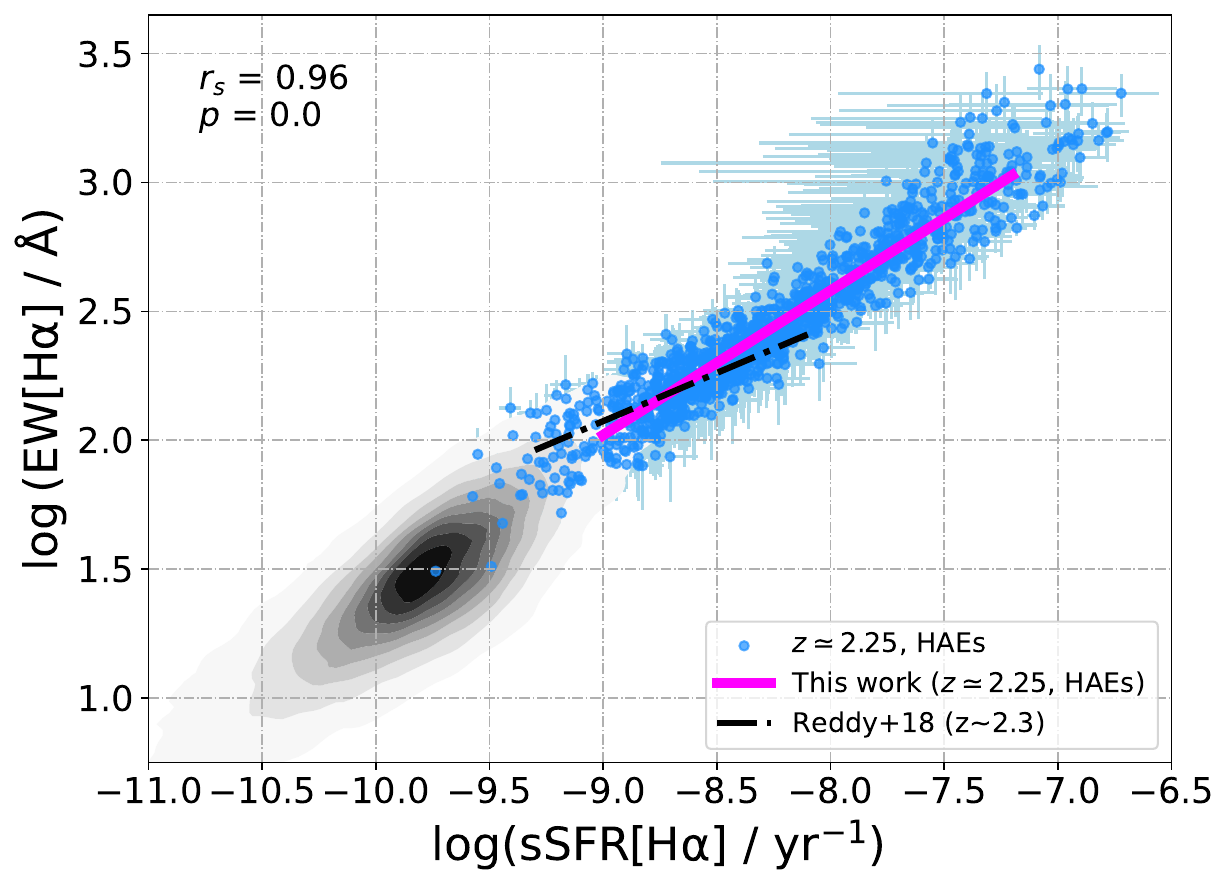}
    \label{fig:ewhavsgal}
    \caption{Relationship between the $EW_\mathrm{{H\alpha}}$ and stellar properties (stellar mass, age, SFR, sSFR) of the 1318 HAEs at $z\sim2.3$ in our study. Outlines as in Figure \hyperref[fig:ewo3vsgal]{7}.}
    \vspace{0.5cm}
\end{figure*}

\begin{table*}[hbt!]
    \small
    \centering
    \label{tab:ewproper}
    \caption{Relationship between the Equivalent Widths and various Galaxy Properties}
    \begin{tabular}{cccccc}
        \hline\hline\hline
        $\qquad$Line$\mathrm{\,^{a}}\qquad$  & Attribute$\mathrm{\,^{a}}$ & $N_{gal}$$\mathrm{\,^{b}}$ &  $\quad\quad$Intercept$\mathrm{\,^{c}}\quad$ & $\quad\quad$Slope$\mathrm{\,^{c}}\quad$ & $\quad\quad\quad r_s$$\mathrm{\,^{d}}\quad\quad$\\
        \hline
        $\mathrm{[O\pnt{III}]}$ & $M_*$/$\,M_{\odot}$ & 859 (All sample) & $7.321\pm0.098$ & $-0.527\pm0.010$ & $-0.864$ \\
        & & 626 (w/ [O{\sc ii}]) & $7.411\pm0.142$ & $-0.536\pm0.015$ & $-0.818$\\
        & & 233 (w/o [O{\sc ii}]) & $7.339\pm0.173$ & $-0.530\pm0.020$ & $-0.886$ \\
        \cline{2-6}
        & Age/$\,$yr  & 859 (All sample) & $9.322\pm0.198$ & $-0.819\pm0.023$ & $-0.762$ \\
        & & 626 (w/ [O{\sc ii}]) & $8.991\pm0.283$ & $-0.786\pm0.033$ & $-0.685$ \\
        & & 233 (w/o [O{\sc ii}]) & $8.703\pm0.231$ & $-0.730\pm0.028$ & $-0.883$  \\
        \cline{2-6}
        & SFR(H$\alpha$)/$\,M_{\odot}\,\mathrm{yr^{-1}}$ & 859 (All sample) & $3.027\pm0.045$ & $-0.550\pm0.038$ & $-0.442$ \\
        & & 626 (w/ [O{\sc ii}]) & $2.782\pm0.054$ & $-0.395\pm0.042$ & $-0.352$ \\
        & & 233 (w/o [O{\sc ii}]) & $3.170\pm0.101$ & $-0.548\pm0.106$ & $-0.202$  \\
        \cline{2-6}
        & sSFR(H$\alpha$)/$\,\mathrm{yr^{-1}}$  & 859 (All sample) & $8.147\pm0.116$ & $\ \ 0.701\pm0.014$ & $\ \ 0.860$ \\
        & & 626 (w/ [O{\sc ii}]) & $8.162\pm0.165$ & $\ \ 0.704\pm0.020$ & $\ \ 0.806$ \\
        & & 233 (w/o [O{\sc ii}]) & $7.630\pm0.172$ & $\ \ 0.629\pm0.022$ & $\ \ 0.887$   \\
        \cline{2-6}
        & O32 & 626 (w/ [O{\sc ii}]) & $2.347\pm0.010$ & $\ \ 1.003\pm0.035$ &  $\ \ \ 0.749$ \vspace{0.2cm}\\ 
        \hline\hline
        $\mathrm{[O\pnt{II}]}$ & $M_*$/$\,M_{\odot}$ & 824 (All sample) & $3.826\pm0.098$ & $-0.187\pm0.010$ & $-0.510$ \\
        & & 626 (w/ [O{\sc iii}]) & $3.965\pm0.109$ & $-0.204\pm0.012$ & $-0.533$ \\
        & & 198 (w/o [O{\sc iii}]) & $3.876\pm0.216$ & $-0.187\pm0.022$ & $-0.482$ \\
        \cline{2-6}
        & Age/$\,$Myr  & 824 (All sample) & $4.081\pm0.167$ & $-0.240\pm0.019$ & $-0.400$ \\
        & & 626 (w/ [O{\sc iii}]) & $4.551\pm0.183$ & $-0.297\pm0.021$ & $-0.465$ \\
        & & 198 (w/o [O{\sc iii}]) & $3.978\pm0.416$ & $-0.221\pm0.047$ & $-0.306$ \\
        \cline{2-6}
        & SFR(H$\alpha$)/$\,M_{\odot}\,\mathrm{yr^{-1}}$ & 824 (All sample) & $2.210\pm0.028$ & $-0.151\pm0.022$ & $-0.237$ \\
        & & 626 (w/ [O{\sc iii}]) & $2.180\pm0.030$ & $-0.132\pm0.024$ & $-0.193$ \\
        & & 198 (w/o [O{\sc iii}]) & $2.315\pm0.066$ & $-0.215\pm0.050$ & $-0.363$  \\
        \cline{2-6}
        & sSFR(H$\alpha$)/$\,\mathrm{yr^{-1}}$  & 824 (All sample) & $4.202\pm0.114$ & $\ \ 0.260\pm0.014$ & $\ \ 0.524$ \\
        & & 626 (w/ [O{\sc iii}]) & $4.328\pm0.123$ & $\ \ 0.278\pm0.015$ & $\ \ 0.574$ \\
        & & 198 (w/o [O{\sc iii}]) & $4.424\pm0.266$ & $\ \ 0.279\pm0.031$ & $\ \ 0.449$ \\
        \cline{2-6}
        & O32 & 626 (w/ [O{\sc iii}]) & $2.014\pm0.008$ & $-0.092\pm0.029$ &  $\ -0.093$ \vspace{0.2cm} \\
        \hline\hline
        $\mathrm{H\alpha}$ & $M_*$/$\,M_{\odot}$ & 1318 (All sample) & $5.867\pm0.057$ & $-0.365\pm0.006$ & $-0.843$ \\
        & & 626 (w/ [O{\sc iii}], [O{\sc ii}]) & $5.477\pm0.106$ & $-0.326\pm0.011$ & $-0.694$ \\
        \cline{2-6}
        & Age/$\,$Myr  & 1318 (All sample) & $7.253\pm0.131$ & $-0.564\pm0.015$ & $-0.707$ \\
        & & 626 (w/ [O{\sc iii}], [O{\sc ii}]) & $6.806\pm0.184$ & $-0.522\pm0.022$ & $-0.629$ \\
        \cline{2-6}
        & SFR(H$\alpha$)/$\,M_{\odot}\,\mathrm{yr^{-1}}$ & 1318 (All sample) & $2.803\pm0.027$ & $-0.321\pm0.023$ & $-0.357$ \\
        & & 626 (w/ [O{\sc iii}], [O{\sc ii}]) & $2.436\pm0.037$ & $-0.056\pm0.030$ & $-0.048$ \\ 
        \cline{2-6} 
        & sSFR(H$\alpha$)/$\,\mathrm{yr^{-1}}$  & 1318 (All sample) & $7.060\pm0.041$ & $\ \ 0.560\pm0.005$ & $\ \ 0.955$ \\
        & & 626 (w/ [O{\sc iii}], [O{\sc ii}]) & $6.826\pm0.060$ & $\ \ 0.536\pm0.007$ & $\ \ 0.937$ \\
        \cline{2-6}
        & O32 & 626 (w/ [O{\sc iii}], [O{\sc ii}])  & $2.383\pm0.009$ & $\ \ 0.324\pm0.033$ &  $\ \ \ 0.365$ \vspace{0.2cm}\\
        \hline\hline
    \end{tabular}
    \begin{tablenotes}
    \item \textbf{Notes.} ${ }^{\text {a }}$ All attributes are calculated as the log scale to exhibit the correlation with log($EW/\,\mathrm{\AA}$) for the line listed leftmost.   ${\quad}^{\text {b }}$ The HAEs in our study are separated into subsample based on whether they have detection of other lines (w/) or not (w/o). ${\quad}^{\text {c }}$ Intercept and slope are obtained from the best-fit linear relationship between the galaxy properties and the equivalent width. ${\quad}^{\text {d }}$ The Spearman's rank correlation coefficient of the galaxy properties and the equivalent width.
    \end{tablenotes}
    \vspace{0.5cm}
\end{table*}

The [O{\sc iii}] equivalent widths as a function of physical parameters (stellar mass, age, SFR, sSFR) are shown in Figure \hyperref[fig:ewo3vsgal]{7}. It should be noted that not all HAEs with [O{\sc iii}] lines have detections of [O{\sc ii}] lines. 
This could be due to uncertainties in flux measurements near the Balmer and 4000$\mathrm{\AA}$ break, where the rest-frame wavelength of the [O{\sc ii}] line is located. Also, galaxies at $z\sim2$ often show intrinsic [OIII]/[OII] line ratios greater than unity \citep[e.g.,][]{Shapley15,Sanders16}. The weaker [O{\sc ii}] line may not reach the detection limit of our medium-band flux excess. For those HAEs with [O{\sc iii}] detections but without [O{\sc ii}] detections, we mark them as open circles in Figure \hyperref[fig:ewo3vsgal]{7} to differentiate them from HAEs with both detections.

The upper-left panel of Figure \hyperref[fig:ewo3vsgal]{7} highlights a clear trend between log$(EW_{\mathrm{[O\pnt{III}]}})$ and $M_*$, confirming a trend that has also been observed in previous studies such as \citet{Reddy18} and \citet{Tang19}. This trend is interpreted as an anti-correlation between equivalent widths and the stellar continuum flux density. In galaxies with higher equivalent widths, the contribution of the stellar continuum to the total flux is reduced. Since the stellar continuum is closely related to stellar mass, it is not surprising that our sample demonstrates such an anti-correlation. When comparing our best-fit results with those from MOSDEF \citep{Reddy18}, we observe a very similar slope and intercept. On the other hand, our study successfully extends this relation to the lower mass domain, around $\sim10^8M_\odot$. This new finding highlights the prevalence of high [O{\sc iii}] equivalent widths in low-mass galaxies at high redshift, which was not previously well-documented.

It is also found that the stellar age obtained from SED fitting has an anti-correlation to log$(EW_{\mathrm{[O\pnt{III}]}})$ in the upper-right panel of Figure \hyperref[fig:ewo3vsgal]{7}. This observation is consistent with tests conducted on photoionization models, which have shown a strong correlation between $EW_\mathrm{{[O\pnt{III}]}}$ and stellar age in starburst events, i.e., simple stellar populations \citep{Stasinska96}. According to these tests, $EW_\mathrm{{[O\pnt{III}]}}$ can decrease by more than two magnitudes within a time span of $10^7\,\mathrm{yr}$, even more than recombination lines. Therefore, $EW_\mathrm{{[O\pnt{III}]}}$ can serve as proxies for the ratio of the current rate of star formation and the past integrated SFR \citep{Reddy18}. The HAEs with the largest $EW_\mathrm{{[O\pnt{III}]}}$  in our sample are likely in the early stage of star formation, with rapidly rising SFRs. This rapid increase in SFRs is accompanied by higher $EW_\mathrm{{[O\pnt{III}]}}$, as we have observed. On the other hand, the star formation history (SFH) of older ($>10^{8.5}\,\mathrm{yr}$) HAEs is more complex, resulting in a wider distribution of $EW_\mathrm{{[O\pnt{III}]}}$ values and the gradual breakdown of the linear relation as the stellar age increases.

In the bottom-left panel of Figure \hyperref[fig:ewo3vsgal]{7}, we examine the relationship between SFR and log$(EW_{\mathrm{[O\pnt{III}]}})$. While previous studies by \citet{Reddy18} and \citet{Tang19} did not find strong variations in SFR with log$(EW_{\mathrm{[O\pnt{III}]}})$, our results suggest a possible anti-correlation between two variables. This discrepancy may be attributed to sample selection biases across different studies. \citet{Reddy18} focused on the MOSDEF sample, which primarily consists of galaxies with stellar masses larger than $\sim10^{9.5}M_\odot$. If we restrict our analysis to galaxies in our sample with $M_*>10^{9.5}M_\odot$, the best-fit result yields a slope of $-$0.06, which is consistent with the one by \citet{Reddy18} (see Appendix \hyperref[sec:sfrew2]{B}). This result is significantly different from the slope obtained when considering the full sample ($-$0.55). The SFMS in Figure \hyperref[fig:sfms]{6} further supports this explanation, as lower-mass HAEs tend to scatter above the SFMS and exhibit higher $EW_\mathrm{{[O\pnt{III}]}}$ values compared to main-sequence galaxies with similar SFR. This discrepancy contributes to the steeper slope observed when fitting the full sample. \citet{Tang19} specifically selected extremely O3Es with $EW_{\mathrm{[O\pnt{III}]}}>200\mathrm{\AA}$, which excludes galaxies with lower EW even in the high-mass domain. Also, the sample size in \citet{Tang19} is smaller compared to ours, potentially leading to biases in their analysis. Although our results demonstrate a discrepancy, the relationship between log$(EW_{\mathrm{[O\pnt{III}]}})$ and SFR is generally less significant compared to other parameter combinations, supported by the Spearman's rank correlation coefficient.

The bottom-right panel of Figure \hyperref[fig:ewo3vsgal]{7} shows the correlation between sSFR and log$(EW_{\mathrm{[O\pnt{III}]}})$. We observe a strong correlation where HAEs with higher sSFRs tend to have higher $EW_\mathrm{{[O\pnt{III}]}}$ values. The [O{\sc iii}] luminosity is known to be an indicator of SFR \citep{Maschietto08}, and the [O{\sc iii}]-calibrated SFR is consistent with the UV-measured SFR for high-redshift emitters \citep[e.g.,][]{Suzuki15}. On the other hand, the continuum luminosity scales with $M_*$. Thus, it is not surprising to observe a good correlation between $EW_\mathrm{{[O\pnt{III}]}}$ and sSFR. Our results further support the notion that sSFR can serve as a useful indicator for $EW_\mathrm{{[O\pnt{III}]}}$ and can potentially be applied even at higher redshifts.
In addition, our sample is located in the extrapolated regions of the sequence derived from the SDSS sample. The discrepancy in the locations of these samples is likely attributed to differences in galaxy properties. At high redshift, the molecular gas fraction become larger \citep{Geach11}. The increasing fraction of molecular gas along with the redshift leads to the evolution of SFMS \citep[e.g.,][]{Speagle2014, Whitaker2014}, that high-redshift galaxies are having larger sSFRs. Because the $EW_\mathrm{{[O\pnt{III}]}}$  also have dependence on sSFR, it will follow the same trend and our sample is also evolving along the vertical axis of the panel.

\subsubsection{$\mathrm{[O\pnt{II}]}$ EWs}
\label{sec:o2gal}
In Figure \hyperref[fig:ewo2vsgal]{8} and Table \hyperref[tab:ewproper]{3}, we present the [O{\sc ii}] equivalent widths as a function of various physical parameters such as stellar mass, age, SFR, and sSFR. Similar to the [O{\sc iii}] equivalent widths, some HAEs have [O{\sc ii}] detections but no [O{\sc iii}] detections, and we distinguish them by marking them as open circles in Figure \hyperref[fig:ewo2vsgal]{8}.

While the [O{\sc ii}] equivalent widths do show correlations with stellar mass, stellar age, and sSFR, these relationships are generally weaker compared to the [O{\sc iii}] equivalent widths, as indicated by the Spearman's rank correlation coefficient. This observation is consistent with the findings of \citet{Reddy18}, who also reported the least significant correlations for the [O{\sc ii}] equivalent width among various emission lines, including [O{\sc iii}], H$\beta$, H$\alpha$.

The weaker dependence of [O{\sc ii}] equivalent widths on these attributes can be attributed to intrinsic differences between [O{\sc ii}] and [O{\sc iii}]. In extreme interstellar medium (ISM) environments, neutral oxygen atoms are more likely to be excited to doubly ionized oxygen ($\mathrm{O^{++}}$) rather than singly ionized oxygen ($\mathrm{O^{+}}$). Consequently, stronger [O{\sc iii}] emisson lines are more commonly observed at higher redshifts. This leads to a significant number of low-mass galaxies exhibiting extremely high [O{\sc iii}] /[O{\sc ii}] ratios due to lower metallicity and higher ionization parameters \citep[e.g.][]{Cardamone09,Erb10, Richard11,Nakajima20}. 
Consistent with this inference, in the upper-left panel of Figure \hyperref[fig:ewo3vsgal]{7}, galaxies with [O{\sc iii}] but without [O{\sc ii}] emission are predominantly low-mass galaxies ($<10^{9}M_\odot$), while in Figure \hyperref[fig:ewo2vsgal]{8}, galaxies with [O{\sc ii}] but without [O{\sc iii}] emission are more concentrated in the massive galaxy regime ($>10^{9.5}M_\odot$).

Nonetheless, when comparing our sample with the SDSS sample, we find that the [O{\sc ii}] equivalent widths of galaxies at $z\sim2.3$ are relatively larger than those of local galaxies. This suggests that the [O{\sc ii}] equivalent widths still exhibit some dependence on the star-formation activities in galaxies, despite the weaker correlations with other physical parameters.

\subsubsection{$\mathrm{H\alpha}$ EWs}
\label{sec:hagal}
The $\mathrm{H\alpha}$ equivalent widths as a function of aforementioned stellar properties are also shown in Figure \hyperref[fig:ewhavsgal]{9} and Table \hyperref[tab:ewproper]{3}. The equivalent widths of recombination lines also work as proxies for stellar mass and stellar age in literature \citep[e.g.,][]{Reddy18,Faisst19,Atek22}. Typically, galaxies with strong [O{\sc iii}] are also characterized by strong recombination lines. Thus, it is not surprising that the correlations between the $\mathrm{H\alpha}$ equivalent width and these physical parameters are found to be similar to those observed for the [O{\sc iii}] equivalent width.

The $\mathrm{H\alpha}$ recombination line directly reflects the ongoing star formation rate (SFR), while stellar mass is closely related to the stellar continuum. Therefore, it is not surprising that log$(EW_{\mathrm{H\alpha}})$ exhibits the strongest dependence on sSFR, as sSFR is a measurement of the current star formation activity relative to the stellar mass. Interestingly, when examining the correlation between the equivalent width of these emission lines and  sSFR for the SDSS sample, i.e., the bottom-right panel of Figure \hyperref[fig:ewo3vsgal]{7}, \hyperref[fig:ewo2vsgal]{8}, \hyperref[fig:ewhavsgal]{9}, we find that the slope of the best-fit linear correlation between these two attributes remains unchanged only for H$\alpha$. On the other hand, for our HAEs sample, the slope of the correlation between sSFR and equivalent widths of [O{\sc iii}] and [O{\sc ii}] are different to that of the SDSS sample. This finding probably contradicts the statement by \citet{Reddy18} that the sSFR-EW relationship is largely redshift-invariant for all emission lines. While our result indicates that this relation is likely to be redshift-invariant only for H$\alpha$ emission. We further discuss this issue in Section \hyperref[sec:JWST]{5.3}. 
\\

Overall, our sample indicates that both the H$\alpha$ and [O{\sc iii}] equivalent widths are sensitive to stellar mass, stellar age, and specific SFRs. Galaxies with larger H$\alpha$ or [O{\sc iii}] equivalent widths tend to have lower stellar masses, younger stellar populations, and higher sSFRs. In contrast, the [O{\sc ii}] equivalent widths show a much weaker dependence on these parameters. Note that, the HAEs in our study represents a less biased sample compared to previous works, which may have biased towards high-mass emitters. Although different selection biases exist, the overall trend does not change significantly.

\subsection{$EW$ vs. ISM properties}
\label{sec:ewism}
The results of Section \hyperref[sec:ewsfr]{4.1} demonstrate the response of the equivalent widths of [O{\sc iii}], [O{\sc ii}], H$\alpha$ to stellar properties. Here, we further discuss their dependence on ISM properties. The ISM properties include gas-phase metallicity, ionization parameters, electron density, which can be indicated by various line indices. Commonly used metallicity-sensitive line indices include N2 ([N{\sc ii}]/H$\alpha$), O3N2 (([O{\sc iii}]/H$\beta$)/([N{\sc ii}]/H$\alpha$)), R23 \citet{Kewley02,Kobulnicky04,Pettini04}. On the other hand, O32 and O3 ([O{\sc iii}]/H$\beta$) are often used as ionization-sensitive line indices. The line ratios of [O{\sc ii}] and [S{\sc ii}] doublets can serve as electron density diagnostics. In our work, we have access to the O32 and R23 line indices. It is important to note that the R23 index does not vary monotonically with gas-phase metallicity, but instead follows an evolutionary path combined with the O32 index, as indicated by photoionization models \citep[e.g.,][]{Kewley02,Ferland13}. We would like to discuss the O32 versus R23-index diagram in Section \hyperref[sec:haelae]{5.1}. As a result, in this section, we will focus on illustrating how the equivalent widths vary with O32, highlighting their dependence on the ionization parameter.

\begin{figure}[hbt!]
    \centering
    \includegraphics[width=0.95\linewidth]{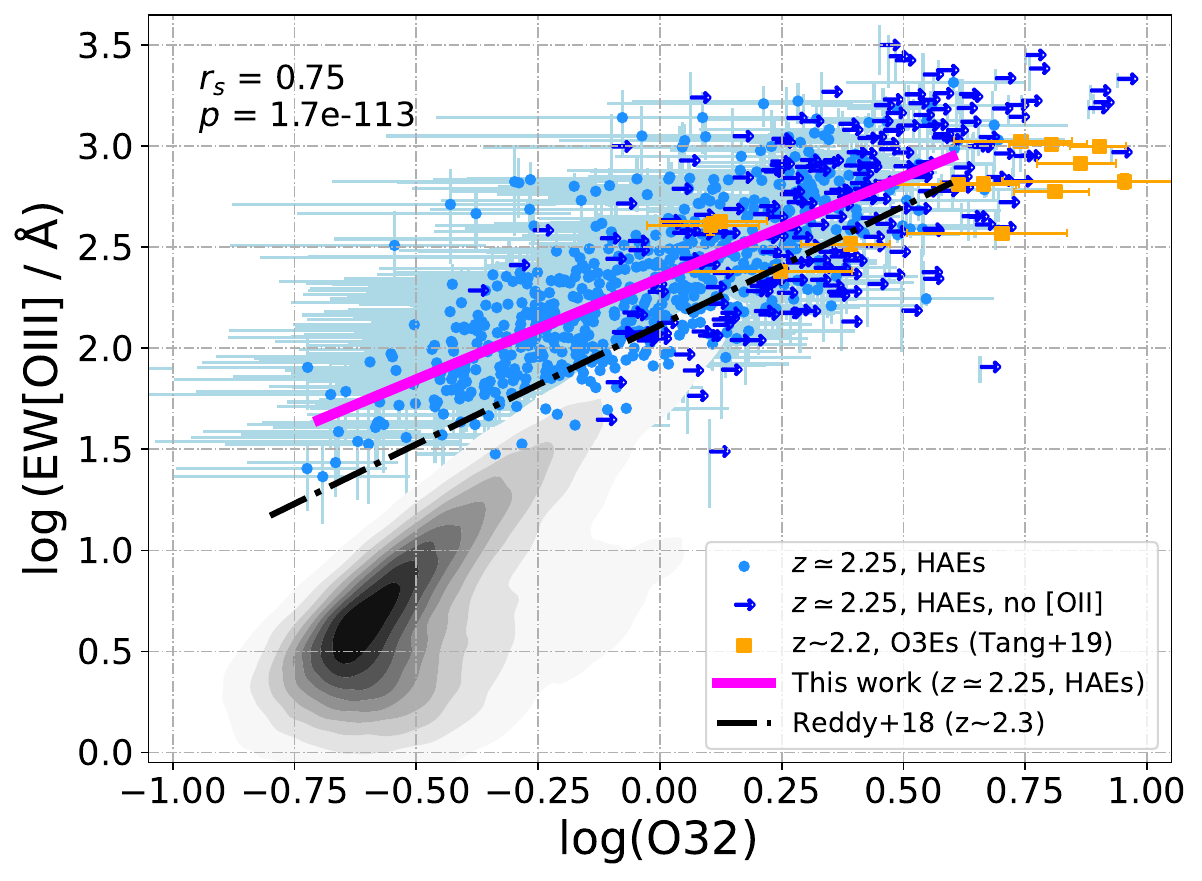}
    \includegraphics[width=0.95\linewidth]{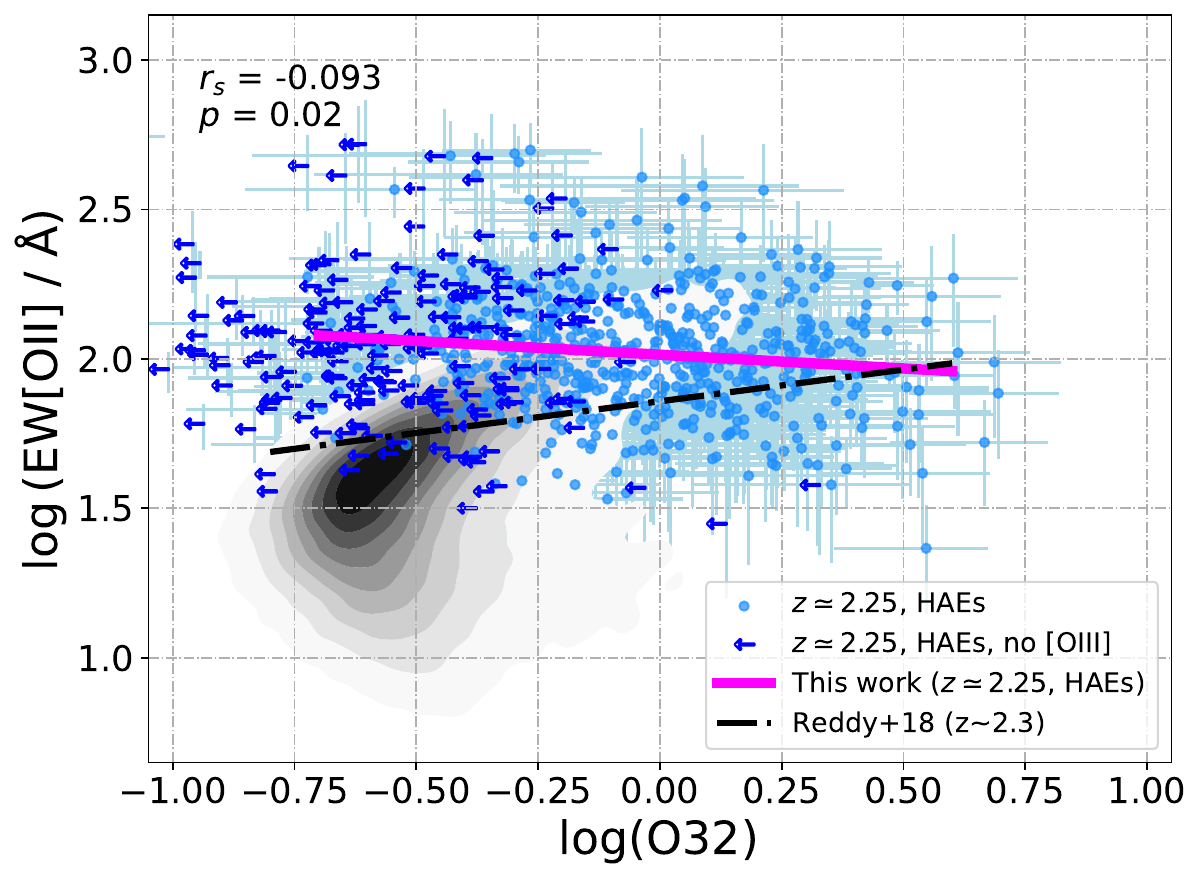}
    \includegraphics[width=0.95\linewidth]{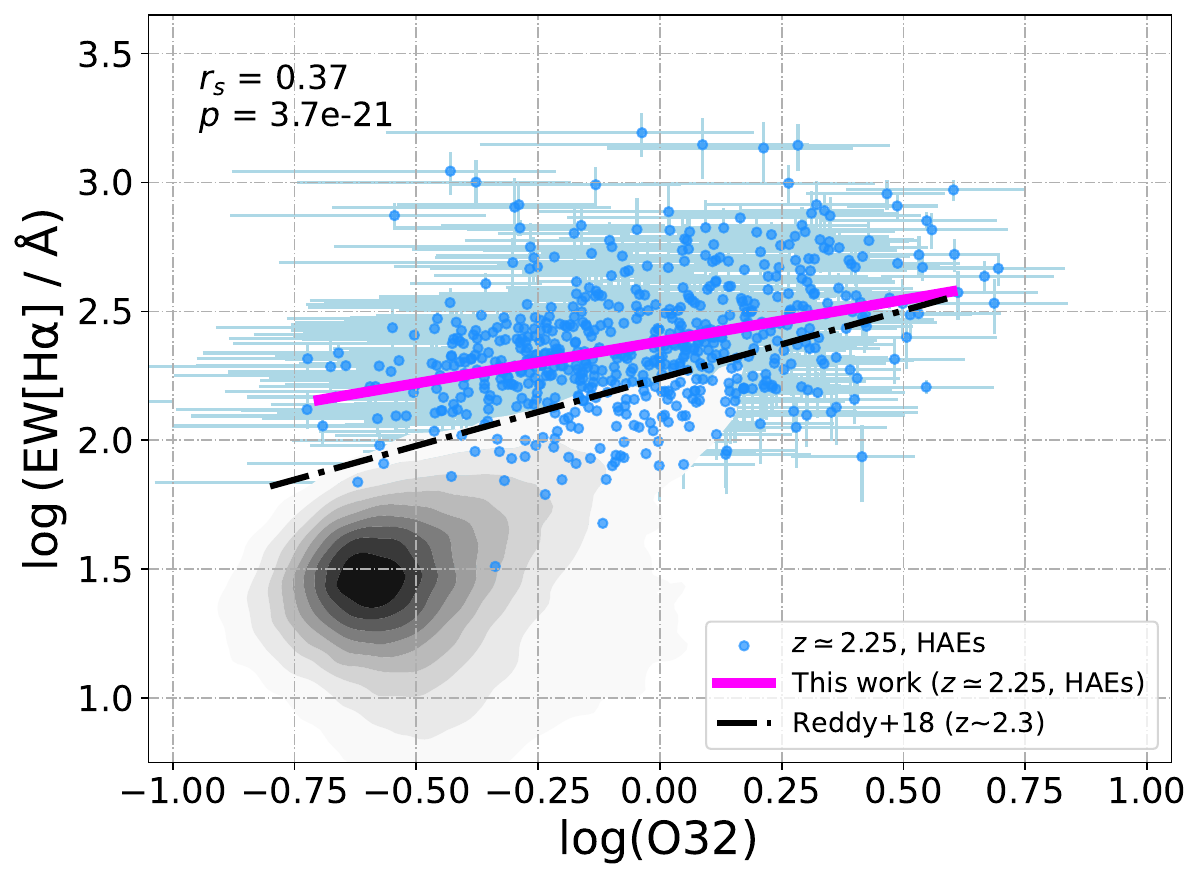}
    \label{fig:ewo32}
    \caption{Relationship between the emission line equvalent widths and ionization-sensitive line index, log(O32), for the HAEs at $z\sim2.3$ in our study. Upper: log$(EW_{\mathrm{[O\pnt{III}]}})$ vs. log(O32). Those HAEs with only [O{\sc iii}] detection are presented by rightward arrows. Middle: log$(EW_{\mathrm{[O\pnt{II}]}})$ vs. log(O32). HAEs with only [O{\sc ii}] detection are presented by leftward arrows. Bottom: log$(EW_{\mathrm{H\alpha}})$ vs. log(O32).}
    \vspace{0.5cm}
\end{figure}

In the top panel of Figure \hyperref[fig:ewo32]{10}, we show the dependence of log$(EW_{\mathrm{[O\pnt{III}]}})$ on the log(O32) index for individual galaxies. For those HAEs without [O{\sc ii}] detections, we assign a lower limit to their O32 values using a 2$\sigma$ upper-limit flux for [O{\sc ii}]. These lower limits are indicated by rightward arrows, and they are not included in the calculation of the best-fit linear correlation presented in Table \hyperref[tab:ewproper]{3}. 

The O32 index serves as a direct indicator of the ionization state of the ISM. We observe a clear trend where O32 increases with increasing $EW_\mathrm{{[O\pnt{III}]}}$, suggesting a harder ionizing radiation field in these high $EW_\mathrm{{[O\pnt{III}]}}$ galaxies. Additionally, O32 exhibits a secondary dependence on gas-phase metallicity, showing an anti-correlation with metallicity \citep[e.g.,][]{Kewley02,Bian18}. This could also imply that these high $EW_\mathrm{{[O\pnt{III}]}}$ galaxies are likely to have lower metallicities.

For HAEs without [O{\sc ii}] detections, we find a median lower limit of $\mathrm{O32}\gtrsim2.2$, which is significantly higher than the median O32 value for the rest of the sample, which is close to unity. Among these non-detections, the subset with the highest lower limits for O32 ($\mathrm{O32}\gtrsim5$) all have $EW_\mathrm{{[O\pnt{III}]}}$ values exceeding $500\mathrm{\AA}$. These galaxies with the highest O32 values, indicative of high ionization parameters in the ISM, are suggested to be powered by extremely young and massive stellar populations.

The middle panel of Figure \hyperref[fig:ewo32]{10} shows how log$(EW_{\mathrm{[O\pnt{II}]}})$ varies with the log(O32) index for our sample. Similarly to the previous panel, we assign a 2$\sigma$ upper-limit flux for [O{\sc iii}] to the objects without [O{\sc iii}] detections, indicating their upper limits for the O32 values (indicated by leftward arrows). These objects are not included in the fitting of the linear correlation presented in Table \hyperref[tab:ewproper]{3}.

In contrast to the strong correlation observed between $EW_\mathrm{{[O\pnt{III}]}}$ and O32, we find that log$(EW_{\mathrm{[O\pnt{II}]}})$ is nearly independent of log(O32) in our sample. As mentioned in Section \hyperref[sec:o2gal]{4.1.2}, the [O{\sc ii}] equivalent width is not sensitive to various galaxy properties due to the lack of $\mathrm{O^{+}}$ ions in extreme ISM environments. This explanation is also applicable to the result observed here, where the O32 index directly indicates the ionization state of the ISM. On the other hand, the result from \citet{Reddy18} indicated a correlation between [O{\sc ii}] and O32, although the dependence was less significant than that of [O{\sc iii}]. Based on the distribution of data points in this panel, we found that the discrepancy between the two results is mainly driven by objects with lower O32 ratios in our sample, as they exhibit higher [O{\sc ii}] equivalent widths compared to the MOSDEF sample in \citet{Reddy18}. We suggest that this discrepancy may be explained by a selection bias towards strong [O{\sc ii}] emitters in our galaxy sample.

The bottom panel of Figure \hyperref[fig:ewo32]{10} is the dependence of log$(EW_{\mathrm{H\alpha}})$ on the log(O32) index. We find that the equivalent widths of H$\alpha$ also increase with O32, although not as significantly as [O{\sc iii}]. 
Since we lack objects with $EW_\mathrm{{H\alpha}}<100\mathrm{\AA}$, the selection bias towards strong emitters still exists in our galaxy sample. This bias might result in a shallower slope compared to that reported in \citet{Reddy18}.

Indeed, the overall trend between the equivalent width and ISM properties observed in our sample is similar to the findings in Section \hyperref[sec:ewsfr]{4.1}. We find that the equivalent widths of [O{\sc iii}] are the most sensitive to the ionization parameter in the ISM, followed by H$\alpha$ emission. On the other hand, the [O{\sc ii}] equivalent widths show almost no dependence on the ionization parameter. This indicates that the [O{\sc iii}] emission line is particularly responsive to the ionization state of the ISM, while [O{\sc ii}] is less affected by these ISM properties. These results highlight the different behaviors of emission lines in response to the ionization conditions within galaxies.

In the upper left panels of Figure \hyperref[fig:ewo3vsgal]{7}, there is an offset in $EW_\mathrm{{[O\pnt{III}]}}$ between our sample and SDSS galaxies at a fixed mass, which may be associated with the increase in SFR with redshift at a given mass \citep[e.g.,][]{Whitaker2014}, and/or the decrease in metallicity with redshift \citep[e.g.,][]{Sanders20}. The former is indicated above, where our sample have higher average SFR than the SDSS sample because of the increasing molecular gas fraction. 
To further explore the cause of this evolution, we look into the correlation between $EW_\mathrm{{[O\pnt{III}]}}$, SFR and O32 in more detail. Following the method in \citet{Reddy18}, we calculate how the relationship between the residuals in $EW_\mathrm{{[O\pnt{III}]}}$ versus residuals in SFR varies in bins of residual O32, which are computed by the deviation of each galaxy’s $EW_\mathrm{{[O\pnt{III}]}}$, $\mathrm{SFR(H\alpha)}$, O32 from the best-fit values at the same stellar mass. The best-fit values of $\mathrm{SFR(H\alpha)}$ are taken from Equation \hyperref[equ:sfms]{10}, and those of $EW_\mathrm{{[O\pnt{III}]}}$ and O32 are calculated from the best-fit linear functions of 626 HAEs that have both [O{\sc iii}] and [O{\sc ii}] emission lines detected. The best-fit results of these linear relations are:
\begin{equation}
    \mathrm{log}(EW_\mathrm{{[O\pnt{III}]}}) = -0.54 \times \log M_* + 7.41.
\end{equation}
\begin{equation}
    \mathrm{log(O32)} = -0.24 \times \log M_* + 2.25.
\end{equation}

\begin{figure}[t]
    \includegraphics[width=1\linewidth]{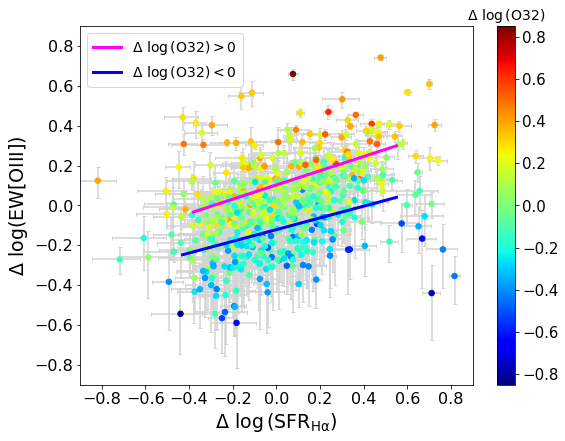}
    \label{fig:sfrcm}
    \vspace{-1.0cm}
    \caption{Residuals of $EW_\mathrm{{[O\pnt{III}]}}$ vs. residuals of $\mathrm{SFR(H\alpha)}$ with the color gradient of dots shows the residuals of O32. The residuals are computed by the deviation of each galaxy’s $EW_\mathrm{{[O\pnt{III}]}}$, $\mathrm{SFR(H\alpha)}$, O32 from the mean values at the same stellar mass. This includes the 626 HAEs with both dection of [O{\sc iii}] and [O{\sc ii}]. The magenta and blue solid lines show the best-fit linear functions with positive and negative residuals of O32, respectively. At a fixed SFR, the equivalent width clearly increases with O32.}
\end{figure}

Figure \hyperref[fig:sfrcm]{11} shows the result. We find that the equivalent width changes with O32 at a fixed offset from the SFMS. As explained above, O32 are primarily correlated with the ionization parameter and also have a anti-correlation to the gas-phase metallicity. Our result proves that the redshift evolution of decreasing metallicity is also a factor to explain the increase in $EW_\mathrm{{[O\pnt{III}]}}$ with redshift at a fixed stellar mass. Physically, the decreasing gas-phase metallicity enable to produce more high-temperature massive stars, which would produce much more ionized photons. These ionized photons are more likely to ionize the surrounding neutral ISM into [O{\sc iii}], leading to the redshift evolution of $EW_\mathrm{{[O\pnt{III}]}}$, which is also observed by \citet{Khostovan16}.

\section{Discussion}
\label{sec:dis}
\subsection{The low-mass HAEs}
From Figure \hyperref[fig:sfms]{6}, we observe that there are a large number of HAEs scattering above the SFMS with a median offset $\Delta\mathrm{MS_{med}}\sim0.3\,\mathrm{dex}$ below the mass completeness limit ($10^{9}M_\odot$).
Such population has also been reported in \citet{Hayashi2016} and \citet{Terao2022}, with high (starburst-like) sSFRs from the strong H$\alpha$ emission. However, more detailed physical properties such as metallicity and ionization parameter of these HAEs have not been studied yet.

In Section \hyperref[sec:result]{4}, we presented the relationships between equivalent widths and various galaxy properties, shedding light on the applications of these emission lines for identifying galaxy properties. To further investigate the aforementioned interesting population, We separate the total 1318 HAEs into two subsets: 401 low-mass ($<10^{9}M_\odot$) HAEs and 917 high-mass ($>10^{9}M_\odot$) HAEs. Among these emission lines examined, the equivalent width of [O{\sc iii}] shows the strongest correlation with most stellar and ISM properties. The low-mass HAEs have an average $EW_\mathrm{{[O\pnt{III}],avg}}\simeq626\mathrm{\AA}$ (taking the $EW$ for the 2$\sigma$ upper-limit fluxes of [O{\sc iii}] for objects with no [O{\sc iii}] detection), whereas the high-mass HAEs have $EW_\mathrm{{[O\pnt{III}],avg}}\simeq145\mathrm{\AA}$.

By applying the relationships in Table \hyperref[tab:ewproper]{3}, the high $EW_\mathrm{{[O\pnt{III}]}}$ observed in these low-mass HAEs indicates the presence of a young stellar population ($<100 \mathrm{Myr}$) and a high ionization state ($\mathrm{O32}\sim3$) in the system. Also, such an elevated $EW_\mathrm{{[O\pnt{III}]}}$ value is typically reported in extreme O3Es studies, like \citet{Yang17} for ``green pea" galaxies in the local universe with  $EW_\mathrm{{[O\pnt{III}],med}}\simeq733\mathrm{\AA}$; \citet{Tang19} for O3Es at $1.3<z<2.4$ with $EW_\mathrm{{[O\pnt{III}],med}}\simeq676\mathrm{\AA}$; and \citet{Onodera20} for O3Es at $3.0<z<3.7$ with $EW_\mathrm{{[O\pnt{III}],med}}\simeq730\mathrm{\AA}$. These O3Es have revealed extreme ionization conditions not commonly seen in older and more massive galaxies, and the low-mass HAEs may also have similar unique properties.

\subsection{Comparison with $z\sim2$ LAEs}
\label{sec:haelae}
Previous multiple emission lines analysis on galaxies at $z\sim2$ mainly focus on massive and bright galaxies, leading to a bias that limits the detailed study of low-mass galaxies at these epochs. 
The inclusion of a large number of low-mass HAEs in our galaxy sample at $z\sim2.3$ is unique compared to previous studies. The high $EW_\mathrm{{[O\pnt{III}]}}$ values observed in these low-mass HAEs further raise interest, as they suggest the possibility of extreme galaxy and ISM properties in this population. Understanding the properties and evolutionary paths of these low-mass galaxies at high redshifts can provide valuable insights into galaxy formation and cosmology.

The relationship between the low-mass HAEs and Ly$\alpha$ emitters (LAEs) is indeed an interesting topic of investigation. LAEs have been identified as potential candidates for Lyman continuum (LyC) leakage, a process thought to be important for cosmic reionization \citep{Robertson13, Bouwens15}. LAEs exhibit specific physical properties, such as high ionization states and low dust absorption, that enable LyC photons to escape into the intergalactic medium (IGM). The association between strong Ly$\alpha$ and [O{\sc iii}] emission lines observed in $z > 7$ LAEs, as reported by \citet{Finkelstein13}, suggests a connection between the extreme [O{\sc iii}] emission and LyC leakage. Additionally, \citet[]{Nakajima14} found a possible correlation between O32 and $f_{esc}$ (the fraction of LyC photons escaping) from LAEs at $z\sim2$, indicating the presence of "density-bounded" H{\sc ii} regions in LAEs that facilitate the escape of ionizing radiation into the IGM.
Considering that a subset of low-mass HAEs in this study have high O32 values, a characteristic often observed in LAEs, it becomes intriguing to explore the relationship between these two populations. Investigating whether low-mass HAEs exhibit characteristics associated with LyC leakage, such as extreme [OIII] emission and low dust absorption, could shed light on their potential role in the cosmic reionization process. Comparing the properties and behaviors of low-mass HAEs and LAEs can provide valuable insights into the diversity and contributions of different galaxy populations.

\begin{figure*}[hbt!]
    \centering
    \includegraphics[width=0.95\textwidth]{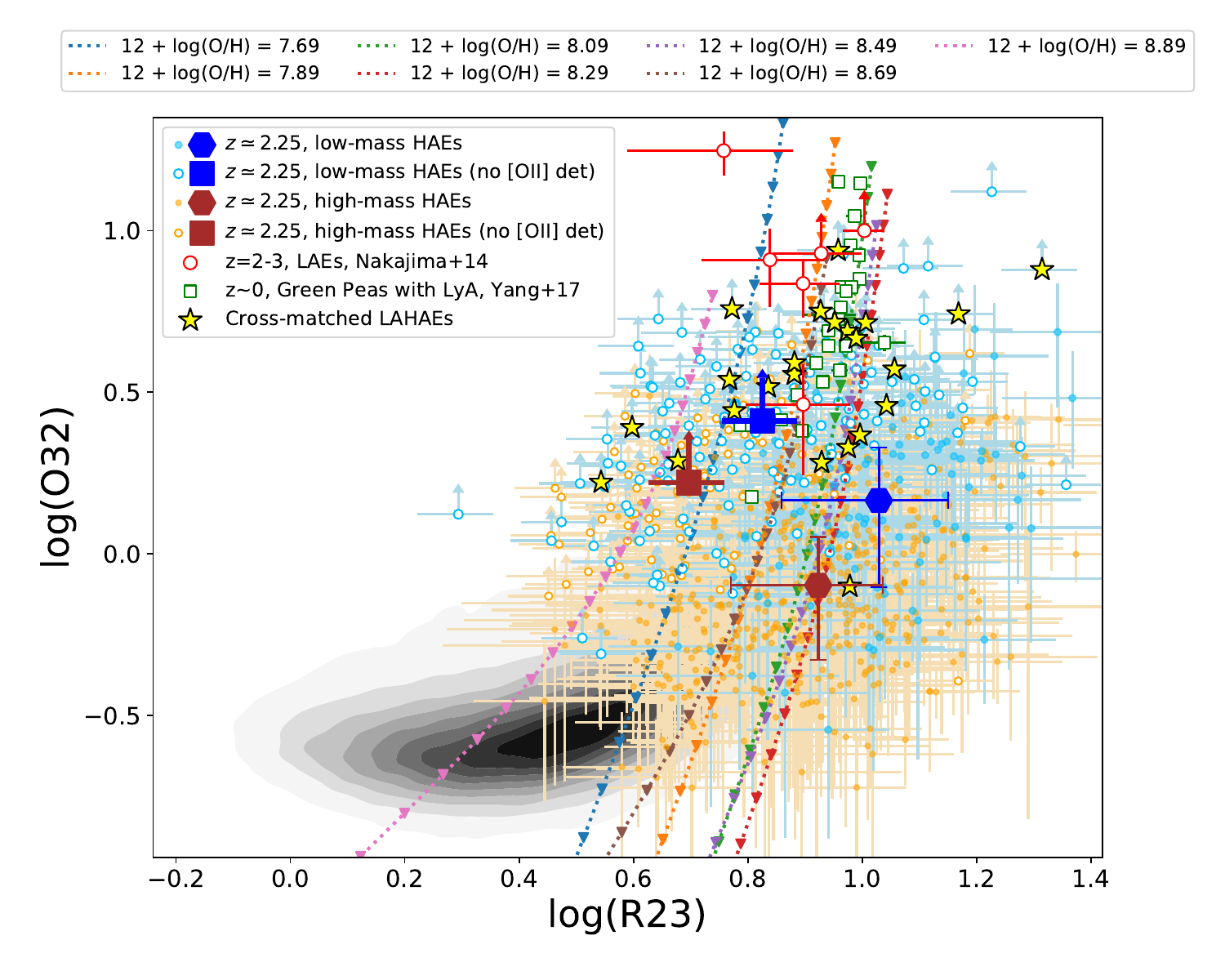}
    \label{fig:o32r23}
    \vspace{-0.5cm}
    \caption{Relation between $\mathrm{O32}$ and $\mathrm{R23}$-index for HAEs at $z\sim2.3$. $\mathrm{[O\pnt{III}]}$ and $\mathrm{[O\pnt{II}]}$ are derived from the flux excesses in $H_s$/$H_l$ and $J_2$/$J_3$ photometry, respectively. Blue (orange) filled circles show the low-mass HAEs (high-mass HAEs) with both detection of $\mathrm{[O\pnt{III}]}$ and $\mathrm{[O\pnt{II}]}$ emission lines, while blue (orange) open circles represent HAEs with only $\mathrm{[O\pnt{III}]}$ detection but no $\mathrm{[O\pnt{II}]}$ detection. Big deeper points are the median stacks of each classification. The 23 cross-matched LAHAEs from \citet{Sobral18} and \citet{Nakajima12} are marked as yellow stars. Red open circles represent the spectroscopic observations of $z \simeq 2–3$ LAEs from \citet{Nakajima14}, and green squares are the ``green pea" with strong Ly$\alpha$ emission from \citet{Yang17}. Local SDSS star-forming/star-burst galaxies are represented by contours. Cloudy+BPASS model emission-line ratios are shown on this O32 vs.R23 diagram. The triangle data points on the curves increase in size as log($U$) increases at fixed nebular metallicity. log($U$) is varied in steps of 0.10 dex from log($U$) = $-$3.6 to $-$1.4 and have nebular metallicity range between 0.1$Z_{\odot}$ to 1.5$Z_{\odot}$. The distribution of HAEs fit well to the model estimation. On the other hand, the low-mass HAEs seem to have higher ionization properties than high-mass HAEs in our study, but lower than those of LAEs.}
    \vspace{0.3cm}
\end{figure*}

We conducted a cross-match between all the galaxies (not only HAEs) at $2.05 < z < 2.5$ in the ZFOURGE catalog (AGN excluded) and two LAEs catalogs: \citet{Nakajima12} (ZF-COSMOS, UDS field) and \citet{Sobral18} (ZF-COSMOS field). \citet{Nakajima12} constructed a catalog of photometric-selected LAEs at $z\sim2.2$ using Subaru narrowband imaging data (NB387). Among their LAE candidates, we found 16 targets in the ZFOURGE-COSMOS field and 19 targets in the ZFOURGE-UDS field, respectively. \citet{Sobral18} created a large sample of $\sim4000$ photometric-selected LAEs at redshifts ranging from $z\sim2$ to 6. They used deep narrow- and medium-band imaging data from the Subaru and Isaac Newton Telescopes in the COSMOS field, covering an area of $\sim2\,\mathrm{deg}^2$. We identified 10 targets in the IA427 and NB392 filters, and none of these targets overlapped with \citet{Nakajima12}. 

In total, we have identified 45 cross-matched galaxies from these two LAEs catalogs. Among them, a subset of 36 cross-matched LAEs shows the detection of [O{\sc iii}] emission lines. In the subsequent discussion, we will mainly focus on these 36 cross-matched LAEs. Besides, within these 36 cross-matched LAEs, 23 LAEs are also identified as HAEs in our study and 19 are classified as the low-mass ($<10^{9}M_\odot$) ones. We refer to the galaxies that hold both Ly$\alpha$ and H$\alpha$ emission as cross-matched LAHAEs.

In Figure \hyperref[fig:o32r23]{12}, we present the O32 versus R23-index diagram, which allows us to investigate the ionization parameter and gas-phase metallicity of galaxies \citep{Kewley02,Kobulnicky04}. The diagram includes the emission line ratios of HAEs at $z\sim2.3$ from our study with star symbols representing the cross-matched LAHAEs. Note that not every sample in our study has simultaneous observations of [O{\sc iii}] and [O{\sc ii}]. Therefore, we only include the samples with [O{\sc iii}] detection in this figure. For those with [O{\sc iii}] detection but no [O{\sc ii}] detection, we also use the 2$\sigma$ upper-limit fluxes of [O{\sc ii}] as lower limits for O32, and they are depicted as open circles with arrows.
To differentiate between the low-mass ($<10^{9}M_\odot$) HAEs and high-mass ($>10^{9}M_\odot$) HAEs, we assign them different colors. Additionally, we plot the median values of these groups as larger and darker symbols for clarity. For comparison, we include spectroscopic observations of $z \simeq 2–3$ LAEs from \citet{Nakajima14}, local ``green pea" galaxies with detection of strong Ly$\alpha$ line ($EW_{\mathrm{Ly\alpha}}>20\,\mathrm{\AA}$) from \citet{Yang17}, as well as local SDSS star-forming and star-burst galaxies. To provide a reference framework, we overlay the O32 versus R23-index curves from the Cloudy \citep{Ferland13} + BPASS \citep{Stanway18} photo-ionization models. Each curve corresponds to a different gas-phase metallicity value ranging from 12+log(O/H)=7.69 to 8.89. The triangles along each curve represent the ionization parameter log($U$), which increase from $-$3.6 (bottom) to $-$1.4 (top). By comparing the observed data with the model curves, we can gain insights into the ionization states and metallicities of the galaxies in our sample.

The median values of our sample demonstrate that low-mass HAEs tend to possess higher ionization parameters compared to high-mass HAEs. Notably, LAEs from \citet{Nakajima14}, \citet{Yang17} and those cross-matched LAHAEs in our study have nearly the highest ionization properties among all the samples. Separately, LAEs in \citet{Nakajima14} have a highest median $\mathrm{O32_{med}}$ of $\sim7.5$; those in \citet{Yang17} have $\mathrm{O32_{med}}\sim5$; while our LAHAEs have $\mathrm{O32_{med}}\sim4$. This trend suggests that LAEs are more likely to exhibit the highest ionization states and potentially leak ionizing photons into the IGM. Although low-mass HAEs generally show lower ionization properties than LAEs, there are several individual objects within this group that display comparably high O32-index values, similar to those observed in LAEs.

Also, the local LAEs from \citet{Yang17} have a median $EW_{\mathrm{H\alpha,med}}\sim500\mathrm{\AA}$ with a median stellar mass of $\sim10^9\,M_{\odot}$, while the LAHAEs in our sample have $EW_{\mathrm{H\alpha,med}}\sim600\mathrm{\AA}$ with a median stellar mass of $\sim10^{8.5}\,M_{\odot}$. These low-z and high-z galaxies hold very close stellar mass, $EW_{\mathrm{H\alpha}}$ and the O32 ratio, demonstrating the similarities in galaxy and ISM properties. This may reveal that the ``green pea" galaxies from \citet{Yang17} are the low-z analogs of the LAHAEs in our study.

\begin{figure}[t]
    \includegraphics[width=1\linewidth]{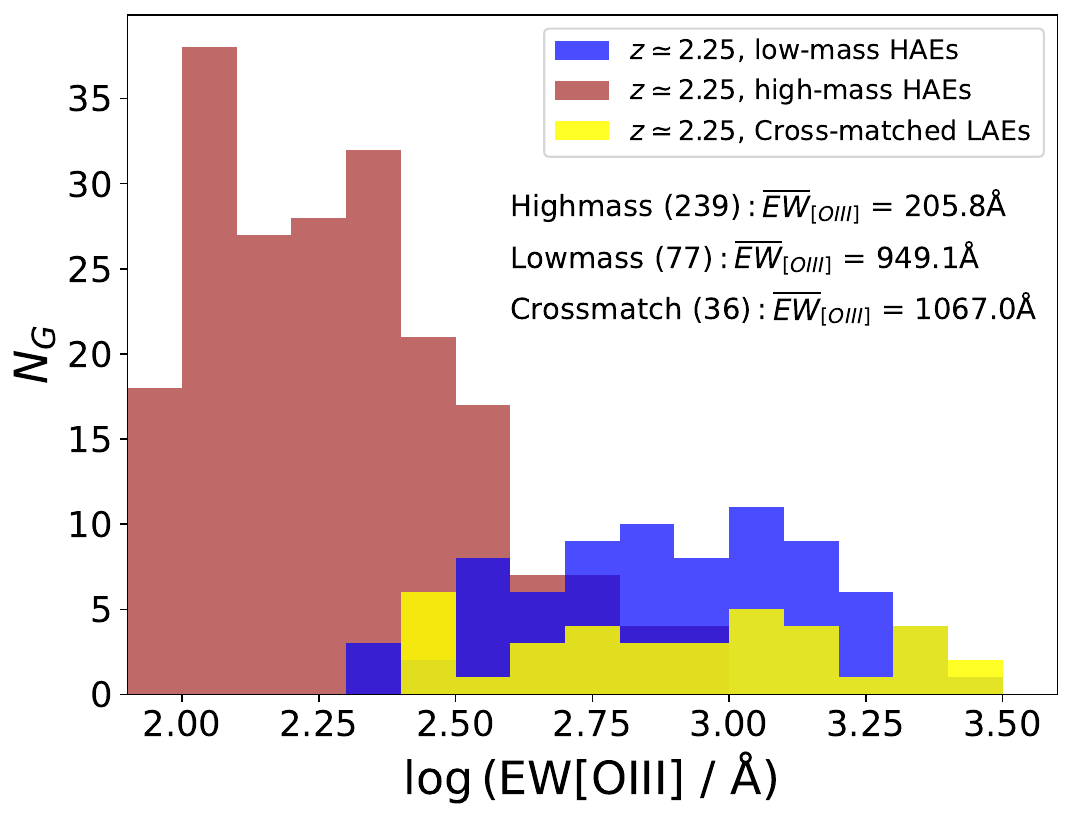}
    \label{fig:laehaeewo3}
    \vspace{-0.5cm}
    \caption{The distribution of the [O{\sc iii}] equivalent width for HAEs in our study, limiting the redshift coverage to $2.14<z<2.26$ and $2.40<z<2.50$. The 77 low-mass HAEs and 239 high-mass HAEs are distributed as a histogram in brown column and blue column. We cross-match $z\sim2.3$ galaxies from the ZFOURGE catalog with two LAEs catalog from \citet{Nakajima12} and \citet{Sobral18} respectively, resulting in a total of 36 cross-matched LAEs with detected [O{\sc iii}] emission lines, represented by the yellow column. The histogram highlights that LAEs tend to have the highest $EW_\mathrm{{[O\pnt{III}]}}$, but are relatively rare as indicated by their significantly lower number density. The high-mass HAEs are the most commonly existing but hold the lowest $EW_\mathrm{{[O\pnt{III}]}}$. The low-mass HAEs occupy an intermediate position between the LAEs and high-mass HAEs, bridging the gap between these two populations.}
\end{figure}

Those galaxies with the highest $EW_\mathrm{{[O\pnt{III}]}}$ values exceeding $1000\mathrm{\AA}$ are likely dominated by extremely hot and massive stars, leading to a more intense radiation field and a higher probability of ionizing photon leakage. In Figure \hyperref[fig:laehaeewo3]{13}, we further investigate the number distribution of $EW_\mathrm{{[O\pnt{III}]}}$ for the high-mass, low-mass HAEs, as well as those 36 cross-matched LAEs. For this analysis, we constrain the redshift coverage of our sample to align with the surveys of \citet{Nakajima12} and \citet{Sobral18}, that is $2.14<z<2.26$ and $2.40<z<2.50$. Interestingly, both the cross-matched LAEs and the low-mass HAEs exhibit significantly higher average $EW_\mathrm{{[O\pnt{III}]}}$ compared to high-mass counterparts. Moreover, the average $EW_\mathrm{{[O\pnt{III}]}}$ values between the cross-matched LAEs and the low-mass HAEs are similar. Though we have identified that 16 cross-matched LAEs have $EW_{\mathrm{[O\pnt{III}]}}>\mathrm{1000\,\AA}$, a larger number of 18 low-mass HAEs without Ly$\alpha$ emission also display $EW_{\mathrm{[O\pnt{III}]}}>\mathrm{1000\,\AA}$.
This suggests the large presence of low-mass HAEs with extremely large [O{\sc iii}] equivalent widths. 
These findings further emphasize the importance of accounting for the contribution of low-mass HAEs, particularly those with high $EW_\mathrm{{[O\pnt{III}]}}$ values, in understanding the mechanisms of cosmic reionization and the escape of ionizing photons.

We further investigate the Ly$\alpha$ detection rate, i.e., the Ly$\alpha$ escape fraction, of the HAEs in our study. Previous studies that examined the dual emitters of Ly$\alpha$ and H$\alpha$ have revealed relatively low Ly$\alpha$ detection rates within their HAEs sample. For instance, \citet{Hayes10b} reported a detection rate of 6 dual emitters out of 55 HAEs, \citet{Oteo15} reported 7 out of 158, and \citet{Matthee16} reported 17 out of 488. Through the cross-matching analysis in this study, we identify 23 cross-matched LAHAEs out of a total sample of 316 HAEs (see in Figure \hyperref[fig:laehaeewo3]{13}), aligning with these earlier findings. However, if we narrow our focus to the low-mass HAEs, a notably higher rate of dual emitters (19/77) emerges. LAEs are typically characterized as blue, low-mass galaxies with low gas-phase metallicity and high star formation rates, which resemble the properties of the low-mass HAEs in our study. Consequently, it is reasonable to interpret the observed higher detection rate of LAHAEs within the low-mass HAEs.

Conversely, we note that approximately half (19/36) of the cross-matched LAEs are classified as the low-mass HAEs, indicating that these two populations still have differences in their properties. One possible explanation is that LAEs and low-mass HAEs reside in different IGM environments, leading to variations in their galaxy populations. \citet{Momose21} have discovered a distinctive characteristic of LAEs, that is their tendency to inhabit regions of higher IGM density while avoiding density peaks. Also, \citet{Shimakawa17b} have found a lack of LAEs in the galaxy cluster cores. Both findings support the depletion of Ly$\alpha$ emission along
the line of sight when penetrating the densest IGM regions. In contrast, other populations such as O3Es and HAEs, do not exhibit this particular preference of IGM environment. Another explanation lies in the selection bias towards strong emitters ($EW_\mathrm{{H\alpha}}>100\mathrm{\AA}$) in our study. LAEs identified in \citet{Nakajima12} are selected down to $EW_\mathrm{{Ly\alpha}}\sim20\mathrm{\AA}$ through narrow-band observation. Those LAEs with modest $EW_\mathrm{{H\alpha}}$ may not be captured with the broad-band technique in this study.

Based on the comparison of galaxy properties, we propose a conceptual ``Iceberg" model to explain the relationship between LAEs and the low-mass HAEs. In this model, we draw an analogy between these two populations and the visible and submerged parts of an iceberg.
LAEs are often regarded as potential LyC leaking candidates due to their extreme properties such as high ionization states and low dust absorption. However, their population is limited, similar to the visible part of an iceberg above the water surface. These LAEs represent a small fraction of the overall galaxy population.
On the other hand, the low-mass HAEs identified in our study can be likened to the submerged part of the iceberg below the water surface. They demonstrate approximately three times higher number density compared to LAEs and generally possess milder ionization states on average. While, it is important to highlight that there exists a significant number of low-mass HAEs that exhibit extreme ISM properties.

\subsection{Implications for JWST observations}
\label{sec:JWST}
In this study, we have utilized photometric data from various ground-based telescopes and the Hubble Space Telescope to identify emitters, including HAEs and O3Es. While, there are still some unresolved issues due to the limitations of the available ZFOURGE dataset. To address these limitations and make further progress, we anticipate that the \emph{James Webb Space Telescope} (\emph{JWST}) data will play a crucial role. The \emph{JWST} offers state-of-the-art spatial resolution and depth, which will allow us to uncover more detailed information about $z\sim2$ HAEs. From the summer of 2022, \emph{JWST} photometric data are gradually released to public \citep[e.g.,][]{Bagley23, Rieke23}. In this section, we discuss the potential advancements that can be achieved with \emph{JWST} data.

As shown in Figure \hyperref[fig:sfmsall]{2}, we exhibit several best-fit SED models for HAEs. Among the panel, the rightmost two black circles are IRAC 1/2 channel. It is evident that for low-mass HAEs (the left two panels), the existing IRAC data are not sufficient to accurately constrain the stellar continuum at longer wavelengths. This is primarily due to the shallower depth of the IRAC observations and the uncertainties caused by source confusion, which is a result of the relatively large PSF size. These limitations in the IRAC data may introduce inaccuracies and reduce the reliability of the SED fitting results. To overcome these limitations, \emph{JWST} observation with the $F277W$, $F444W$ and $F770W$ filters would significantly improve the accuracy of SED fitting and enable more precise measurements of H$\alpha$ luminosities. By combining the \emph{JWST} data with the existing dataset, we can achieve a more robust determination of the stellar continuum and subsequently obtain more accurate measurements of emission lines by accurately accounting for flux excess.

Through our selection criterion, we have identified 401 low-mass HAEs scattering above the SFMS in Figure \hyperref[fig:sfms]{6}. Since the equivalent width of H$\alpha$ hold a very tight correlation to sSFR, the excess on the SFMS is responsible for the observed anti-correlation between between EW(H$\alpha$) and stellar mass at $<10^{9}M_\odot$, as clearly shown in the upper left panel of Figure \hyperref[fig:ewhavsgal]{9}. However, for galaxies with stellar mass larger than $10^{9.5}M_\odot$, the correlation between these two attributes becomes much weaker, indicating that these massive galaxies align well with the SFMS. This raises a question regarding whether this observation reflects the intrinsic scatter of the SFMS or the existence of a low-mass sequence in the H$\alpha$ indicator. A similar feature in the SFMS of HAEs at $z\sim1$ was observed by \citet{Atek22}. They suggested that the excess of SFMS observed in low-mass galaxies is likely a result of H$\alpha$ flux incompleteness. On the other hand, \citet{Terao2022} conducted simulations of mock HAEs, accounting for photometric errors and fluctuations, and plotted them on the SFMS. Their results indicated that high photometric scatter is unlikely to be the cause of the high SFRs observed in low-mass galaxies. At a higher redshift ($z\sim4.5$), \citet{Faisst19} also found that more than half of their galaxies exhibited an excess in H$\alpha$ SFR, attributed to recent increases in their star formation activities. To further investigate this issue, much deeper \emph{JWST} data would be instrumental in extending the H$\alpha$ SFMS to lower-mass completeness, which has not been explored previously.

The higher sensitivity of \emph{JWST} data enables the detection of [O{\sc iii}] and [O{\sc ii}] emission lines more effectively. In Section \hyperref[sec:hagal]{4.1.3}, we mentioned the sSFR-$EW$ relationship is not exactly redshift-invariant for [O{\sc iii}] and [O{\sc ii}]. Our sample predominantly consists of galaxies with $EW_\mathrm{{[O\pnt{III}]}}>30\mathrm{\AA}$, while the SDSS sample reaches the detection limit at several angstroms. Due to the $EW$ incompleteness of our sample, it is difficult to determine whether low $EW$ galaxies would follow the same slope as our high $EW$ sample or align with the SDSS sample region. Our initial assumption leans towards the former. As discussed in Section \hyperref[sec:o2gal]{4.1.2}, higher $EW_\mathrm{{[O\pnt{III}]}}$ is more likely to be observed at higher redshift. Also, from the bottom-left panel of Figure \hyperref[fig:ewo2vsgal]{8}, the SDSS sample exhibit a comparable $EW_\mathrm{{[O\pnt{II}]}}$ to our sample but with a much lower sSFR. These findings suggest that for galaxies with similar sSFR, they may have lower [O{\sc iii}] but higher [O{\sc ii}] in the local universe. In any case, in order to further investigate the accuracy of our statements, following measurements with high sensitivity are indispensable.

\section{CONCLUSIONS}
\label{sec:conclu}
In this work, we have carried out a systematic search for HAEs at $z \sim 2.3$ in three ZFOURGE fields. The selection process for identifying HAE candidates involved examining the flux excess detected in the ZFOURGE-$K_s$ filters, which is indicative of strong H$\alpha$ emission lines compared to the underlying stellar continuum estimated through SED fitting. To ensure the reliability of our sample, we applied a conservative selection criterion of $3\sigma$, resulting in the identification of 1318 HAEs.
\begin{itemize}[leftmargin=*]
    \item We have identified more than 1300 HAEs in three ZFOURGE fields. Considering the limiting volume of the ZFOURGE survey ($\Delta V=6.8\times10^5\,\mathrm{Mpc^3}$), our method has proven to be highly efficient in identifying emitters. Additionally, the derived emission line fluxes, including $\mathrm{H\alpha}$, [O{\sc iii}] and [O{\sc ii}], exhibit a high level of consistency with measurements obtained through spectroscopy (and grism) from the MOSDEF (and 3D-HST) Emission-Line Catalogs. Specifically, more than 80\% of the detected fluxes show consistent values within a factor of 2. This demonstrates the reliability and accuracy of our method in determining the emission line properties.
    \item The $\mathrm{SFR}-M_*$ relation, i.e., SFMS, derived from the $\mathrm{H\alpha}$ luminosity, exhibits a slope of $0.56\pm0.03$ above the stellar mass completeness. When comparing our results with those from the literature, we find that the shallower slope is primarily influenced by sample selection biases. Meanwhile, we identify a subset of 401 low-mass HAEs ($<10^9\ M_{\odot}$) that deviate from the SFMS($\mathrm{H\alpha}$) by $\sim$0.3 dex.
\end{itemize}

Following the same strategy of extracting $\mathrm{H\alpha}$ emission lines, we also extract the [O{\sc iii}] and [O{\sc ii}] emission lines of these HAEs from the ZFOURGE medium $J/H$-band filters. This allows us to investigate the correlations between the equivalent widths of these lines and various galaxy and ISM properties, including stellar mass, stellar age, SFR, sSFR, and ionization state. The main findings of our analysis can be summarized as follows. These findings shed light on the relationships between emission line equivalent widths and galaxy/ISM properties, providing insights into the nature of intense emitters.
\begin{itemize}[leftmargin=*]

    \item The stellar mass, stellar age, sSFR, exhibit significant correlations with the [O{\sc iii}] and H$\alpha$ equivalent widths. These properties serve as useful indicators for identifying intense emitters. Our method successfully extends these correlations to the lower mass domain, around $\sim10^8M_\odot$, supporting the notion that high $EW_\mathrm{{[O\pnt{III}]}}$ and $EW_\mathrm{{H\alpha}}$ are prevalent in low-mass galaxies at high redshift.
    
    \item The [O{\sc ii}] equivalent widths display the weakest correlations with the aforementioned attributes. This suggests that neutral oxygen is more likely to be doubly ionized at higher redshifts. This observation aligns with the presence of a considerable number of low-mass galaxies exhibiting weak [O{\sc ii}] emission at high redshift.

    \item Our sample reveals that the ionization-sensitive line index, O32, increases with the [O{\sc iii}] equivalent widths, indicating extreme ISM properties for the most intense [O{\sc iii}] emitters. In contrast, the H$\alpha$ equivalent widths show a much weaker correlation with ionization states, and the [O{\sc ii}] equivalent widths are largely independent of O32. This implies that optical emission lines have varying sensitivities to ISM properties.
\end{itemize}

Finally, we have compared the galaxy properties of the low-mass HAEs in our study with those of Ly$\alpha$ emitters (LAEs), which are known to have high ionization parameters and are considered as potential LyC leakages. While the low-mass HAEs exhibit milder ionization states on average, a considerable number of them still possess extreme ISM properties. We propose an ``Iceberg" model to connect LAEs and low-mass HAEs, highlighting the importance of low-mass HAEs during cosmic reionization.

To further advance our understanding, future analysis using the extensive data from \emph{JWST} will provide more robust constraints on our study. Also, the deeper and longer wavelength data from \emph{JWST} will enable the construction of a larger sample of low-mass HAEs at higher redshifts, potentially extending down to even lower masses. These observations will significantly enhance our knowledge of the properties and roles of low-mass galaxies on galaxy evolution and cosmic reionization.

\vspace{5mm}

The authors thank referee for the helpful suggestions and comments.
We are grateful for enlightening conversations with Jonathan Cohn, Ken-ichi Tadaki, and Kazuki Daikuhara. 
We thank the MOSDEF collaboration and 3D-HST collaboration for providing their data releases, which include the measurements of redshift and flux for our sample. We also thank the MPA/JHU team for making their measured quantities on SDSS galaxies publicly available. N.C. gratefully thanks Mengtao Tang for providing the O3Es data shown in this work.
We also gratefully thank Naveen A. Reddy for the best-fit relation results in the previous work.
We appreciate the ZFOURGE data products and support from the FourStar instrumentation team and staff at LCO. We are grateful to the Subaru Telescope team members for their support for SWIMS during the commissioning observation. This research is based in part on data collected at the Subaru Telescope, which is operated by the National Astronomical Observatory of Japan. We are honored and grateful for the opportunity of observing the Universe from Maunakea, which has the cultural, historical and natural significance in Hawaii.
Data analysis was in part carried out on the Multi-wavelength Data Analysis System operated by the Astronomy Data Center (ADC), National Astronomical Observatory of Japan.
This work was supported by FY2022 Graduate Research Abroad in Science Program (GRASP) run by the Graduate School of Science at the University of Tokyo and MEXT/JSPS KAKENHI grant Nos. 15H02062, 24244015, 18H03717, 20H00171, 22K21349.

\vspace{5mm}
\facilities{Magellan (FourStar), Subaru (SWIMS)}
\vspace{10mm}

\section*{\textbf{Appendix A\\SWIMS imaging}}
\label{sec:swimsimageapx}
\begin{figure}[hbt!]
    \centering
    \includegraphics[width=1\linewidth]{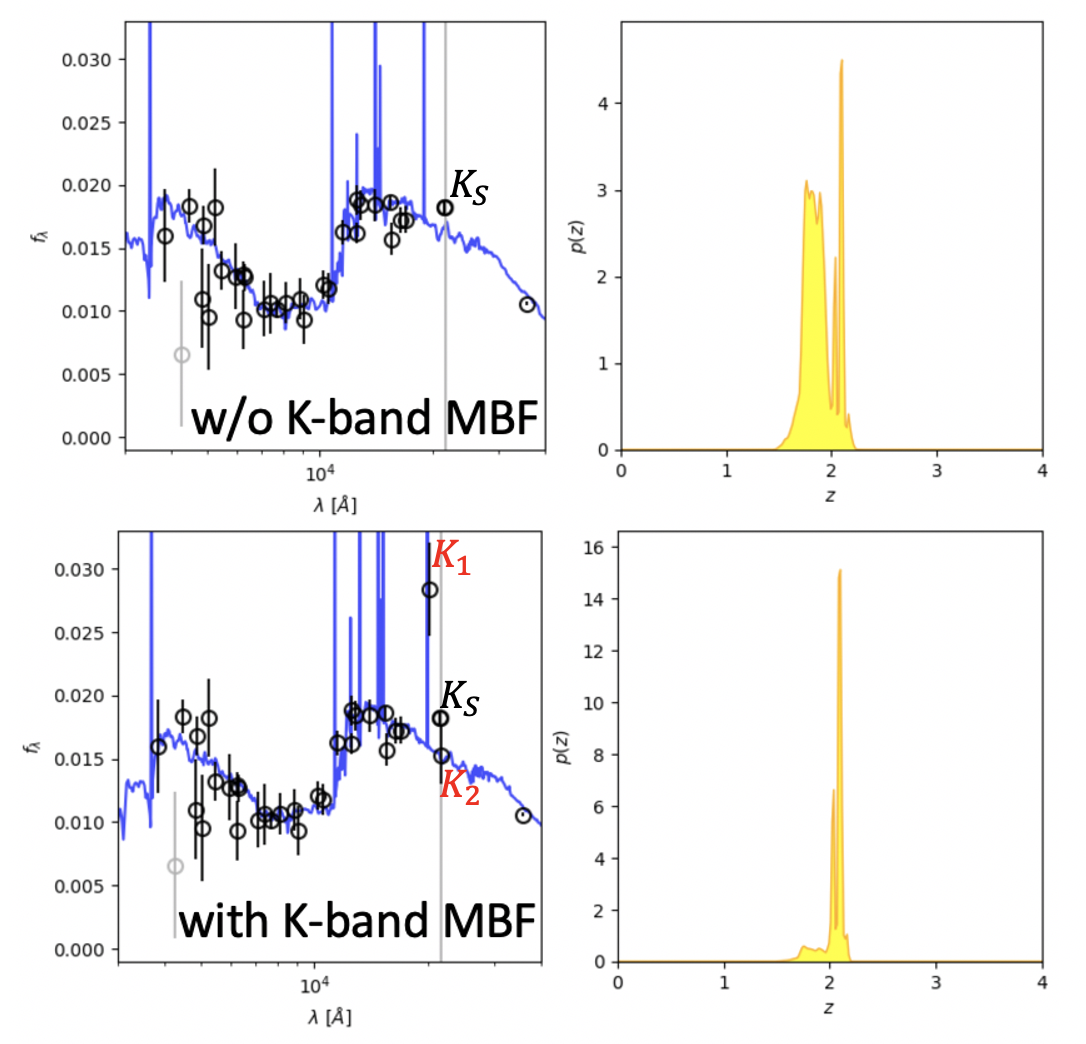}
    \label{fig:eazysamp}
    \vspace{-0.4cm}
    \caption{An example galaxy fitted by EAZY templates to obtain $z_{phot}$. Open circles represent flux of the galaxy in every filter. In the upper panel we show the EAZY result without the SWIMS $K_1$/$K_2$ filters. Blue line represents the best-fit template spectrum. In the bottom panel we show the result after including the SWIMS $K_1$/$K_2$ bands. In both cases, we exhibit the redshift probability distribution functions $p(z)$ in the right panels. A much better constraint is obtained after including the SWIMS $K_1$/$K_2$ data.}
\end{figure}

Figure \hyperref[fig:eazysamp]{14} is an example that SWIMS $K_1$/$K_2$ fluxes is included in the ZFOUREG catalog. We rerun the EAZY code and update $z_{phot}$ (\robotoThin{z\_peak}) with the new outputs. As this object does not show strong color excess in medium $J/H$ band, it shows a bimodal distribution in $p(z)$ when EAZY was run without the SWIMS $K_1$/$K_2$ data. After SWIMS data being included, the $K_1$ filter shows strong color excess, likely to be boosted by H$\alpha$ emission line, while the $K_2$ filter may indicate the level of stellar continuum. With the additional SWIMS data, EAZY no longer gives a bimodal distribution but a very constrained distribution of $p(z)$.

We again obtain the $\sigma_z =|z_{phot}-z_{spec}|/(1+z_{spec})$ as ZFOURGE was done. After including our SWIMS MB data, $\sigma_z$ drop from 0.03 to 0.02 in the ZFOURGE-COSMOS field. Statistically, the overall correspondence is ever better after adding SWIMS $K_1$/$K_2$ data, as indicated by the smaller scatter in the difference between photometric and spectroscopic redshifts.

\section*{\textbf{Appendix B\\SFR vs. $\mathbf{EW_{[O\pnt{III}]}}$ of massive HAEs}}
\label{sec:sfrew2}
\begin{figure}[hbt!]
    \centering
    \includegraphics[width=1\linewidth]{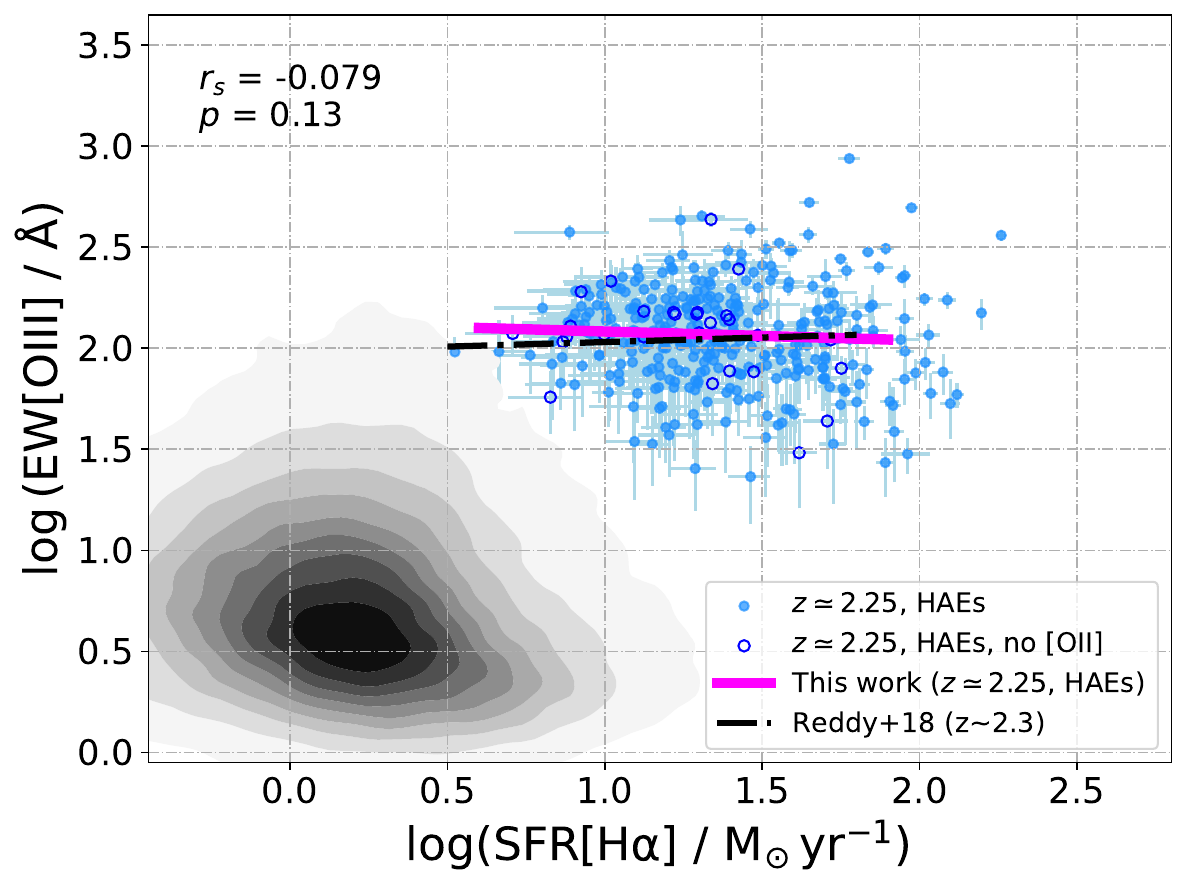}
    \label{fig:sfrew2}
    \vspace{-0.4cm}
    \caption{Relationship between the $EW_\mathrm{{[O\pnt{III}]}}$ and SFR of the HAEs with stellar masses larger than $10^{9.5}M_\odot$. Outlines as in Figure \hyperref[fig:ewo3vsgal]{7}.}
\end{figure}

In this analysis, we have focused on galaxies in our sample with stellar masses greater than $10^{9.5}M_\odot$. This specific subset of galaxies allows us to obtain the best-fit result with a slope of $-0.06$ in Figure \hyperref[fig:sfrew2]{15}. We acknowledge that there is a discrepancy between this specific result and the bottom-left panel of Figure \hyperref[fig:ewo3vsgal]{7}, which could potentially be attributed to sample selection biases.

It is important to consider that sample selection biases can introduce uncertainties and limitations to our findings. Factors such as the selection criteria, observational constraints, and data quality can influence the observed trends and correlations. Therefore, it is crucial to carefully assess and account for any potential biases when interpreting and comparing different results within the analysis.

\section*{\textbf{ORCID iDs}}
\noindent
Nuo Chen \orcidlink{0000-0002-0486-5242}\url{https://orcid.org/0000-0002-0486-5242}\\
Kentaro Motohara \orcidlink{0000-0002-0724-9146}\url{https://orcid.org/0000-0002-0724-9146}\\
Lee Spitler \orcidlink{0000-0001-5185-9876}\url{https://orcid.org/0000-0001-5185-9876}\\
Kimihiko Nakajima \orcidlink{0000-0003-2965-5070}\url{https://orcid.org/0000-0003-2965-5070}\\
Rieko Momose \orcidlink{0000-0002-8857-2905}\url{https://orcid.org/0000-0002-8857-2905}\\
Tadayuki Kodama \orcidlink{0000-0002-2993-1576}\url{https://orcid.org/0000-0002-2993-1576}
Masahiro Konishi \orcidlink{0000-0003-4907-1734}\url{https://orcid.org/0000-0003-4907-1734}

\bibliography{main}{}
\bibliographystyle{aasjournal}

\end{document}